%% file: extended-version.tex
\documentclass[twocolumn]{autart}    

\usepackage{graphicx}      

\usepackage{amsmath} 
\usepackage{amssymb}  
\usepackage{bm}
\usepackage{mathtools}
\usepackage{tabularx}
\usepackage{booktabs}
\usepackage{multirow}
\usepackage{units}
\usepackage{hyperref}
\usepackage{tikz}
\usetikzlibrary{patterns}
\usepackage{pgfplots}
\usepackage{tikzscale}
\pgfplotsset{compat = 1.3}

\usepackage{algorithm}
\usepackage[noend]{algpseudocode}

\newcommand{\diag}{\mathrm{diag}}

\newtheorem{assumption}{Assumption}
\newtheorem{remark}{Remark}
\newtheorem{definition}{Definition}
\newtheorem{problem}{Problem}

\newcommand\mydots{\makebox[1em][c]{.\hfil.\hfil.}}
\newcommand\myvdots{\vspace*{1pt}\vbox{\baselineskip=3.2pt \lineskiplimit=0pt 
\kern7pt \hbox{.}\hbox{.}\hbox{.}}}
\newcommand\myddots{\mathinner{\mkern0mu\raise7pt\vbox{\kern5pt\hbox{.}}\mkern0mu\raise4pt\hbox{.}\mkern0mu\raise1pt\hbox{.}\mkern0mu}}

\begin{document}
\begin{frontmatter}

\title{Computationally Efficient System Level Tube-MPC for Uncertain Systems -- Extended Version}

\author[First]{Jerome Sieber}\ead{jsieber@ethz.ch},
\author[First]{Alexandre Didier}\ead{adidier@ethz.ch},
\author[First]{Melanie N. Zeilinger}\ead{mzeilinger@ethz.ch},

\address[First]{Institute for Dynamic Systems and Control, ETH Zurich, 8092 Zurich, Switzerland}


\begin{abstract}                
Tube-based model predictive control~(MPC) is one of the principal robust control techniques for constrained linear systems affected by additive disturbances. While tube-based methods with online-computed tubes have been successfully applied to systems with additive disturbances, their application to systems affected by additional model uncertainties is challenging. This paper proposes a tube-based MPC method - named filter-based system level tube-MPC (SLTMPC) - which overapproximates both types of uncertainties with an online optimized disturbance set, while simultaneously computing the tube controller online. For the first time, we provide rigorous closed-loop guarantees for receding horizon control of such a MPC method. These guarantees are obtained by virtue of a new terminal controller design and an online optimized terminal set. To reduce the computational complexity of the proposed method, we additionally introduce an asynchronous computation scheme that separates the optimization of the tube controller and the nominal trajectory. Finally, we provide a comprehensive numerical evaluation of the proposed methods to demonstrate their effectiveness.
\end{abstract}

\end{frontmatter}

\section{Introduction}
Robust model predictive control~(MPC) is one of the main robust control techniques for constrained linear systems affected by bounded uncertainties. In the presence of model uncertainty, formulating a robust MPC problem is challenging, since the predicted trajectories are affected by the uncertain model and the computed control inputs, coupling the model uncertainty and the optimization variables. Early approaches to tackle this problem were based on LMI methods~\cite{Kothare1996}, but more recently tube-based MPC methods have become popular for these types of uncertain systems. Tube-based MPC is based on separating the system behavior into nominal and error dynamics, which describe the system dynamics neglecting uncertainties and the deviations from the nominal dynamics due to uncertainties, respectively. Since the uncertainties are assumed to be bounded, the error dynamics can be bounded in so-called tubes, which are computed based on the uncertainty bounds and a tube controller~\cite{Rawlings2009}. However, the challenge of disentangling the model uncertainty and the optimization variables remains, which has been addressed by mainly two different strategies in the literature. One strategy for managing both additive disturbances and model uncertainties with tube-based MPC is lumping both uncertainties into a single state- and input-dependent additive term, which is subsequently bounded offline, resulting in a system with only additive uncertainty. This approach enables the use of various well-established tube-based MPC methods. Standard tube-MPC computes the tubes offline and optimizes the nominal trajectory subject to tightened constraints online, see e.g.~\cite{Chisci2001,Mayne2005,Zanon2021}. To reduce conservativeness, homothetic tube-MPC~\cite{Rakovic2012b} and elastic tube-MPC~\cite{Rakovic2016} enable online optimization of the tubes by fixing the tube shape, but allowing dilation and translation of the tubes online. As shown in~\cite{Sieber2022}, online optimization of the tube controller enables tube flexibility beyond dilation and translation. The online optimization is facilitated by parameterizing the tube controller as a disturbance feedback policy~\cite{Lofberg2003,Goulart2006} or via the system level parameterization~(SLP)~\cite{Sieber2021}, which is part of the system level synthesis framework~\cite{Anderson2019}. Beyond tube-based MPC, the lumped uncertainty approach is also used in classical disturbance feedback control~\cite{Bujarbaruah2021,Bujarbaruah2022}, which tightens the constraints with offline computed bounds as a function of the online computed control input. However, the lumped uncertainty approach is typically conservative, since it relies on offline computed global uncertainty bounds.

Another strategy treats the uncertainties separately and explicitly designs the tubes and the tube controller for the individual uncertainty descriptions. Most approaches following such a strategy build on either the homothetic tube-MPC formulation~\cite{Langson2004,Lorenzen2019,Kohler2019} or the elastic tube-MPC formulation~\cite{Fleming2014,Lu2019}. Additionally, the methods proposed in~\cite{Lorenzen2019,Kohler2019,Lu2019} incorporate online model updates based on set-membership estimation~\cite{Milanese1991} to reduce conservativeness. Other tube-based methods use configuration-constrained tubes~\cite{Villanueva2022}, which do not explicitly parameterize a tube controller but compute a control input for every vertex of the tube, or use ellipsoidal tubes~\cite{Parsi2022}, which are computed via a LFT-based formulation of the uncertainties. More recently,~\cite{Chen2023} proposed a tube-based MPC method based on the SLP that simultaneously overapproximates both uncertainties and computes a tube controller online, which we denote as \emph{filter-based system level tube-MPC~(SLTMPC)}. This method offers highly flexible tubes and requires minimal offline design; however, it provides closed-loop guarantees only for a shrinking horizon formulation and, like all SLP-based MPC approaches, is computationally expensive. The computational issue has been addressed on an algorithmic level by structure-exploiting solvers~\cite{Leeman2024}, and on a system theoretic level by asynchronous computation schemes~\cite{Sieber2023}, which split the tube and nominal trajectory computations into separate processes.

\textit{Contributions:}
The main theoretical contribution of this paper is three-fold: (i) we extend the filter-based SLTMPC method proposed in~\cite{Chen2023} to general polytopic disturbance sets, which results in higher flexibility in the design and provides a more intuitive explanation of the method; (ii) we propose a new terminal controller design and an online optimized terminal set that guarantee recursive feasibility in receding horizon for filter-based SLTMPC; and (iii) we propose a new asynchronous computation scheme that reduces the computational complexity of filter-based SLTMPC methods by splitting the tube and nominal trajectory optimizations into different processes. Additionally, we provide a comprehensive numerical evaluation of the proposed methods on a double integrator and a vertical take-off and landing~(VTOL) vehicle.

The paper is organized as follows: Section~\ref{sec:preliminaries} introduces the notation, the problem formulation, and some basic definitions. Section~\ref{sec:SLTMPC} introduces filter-based SLTMPC generalized from~\cite{Chen2023}, before Section~\ref{sec:SLTMPC-rec-feas} introduces the new terminal ingredients and proves recursive feasibility. Section~\ref{sec:asynch-up} details the asynchronous computation scheme and Section~\ref{sec:numerical_section} showcases the proposed filter-based SLTMPC variants on two numerical examples. Finally, Section~\ref{sec:conclusions} concludes the paper and the Appendix provides details on the implementation of the proposed MPC methods.

The code for the numerical examples is available at \href{https://git.sieber.io/sltmpc}{git.sieber.io/sltmpc}.

\section{Preliminaries}\label{sec:preliminaries} \vspace{-0.15cm}
\subsection{Notation \& Definitions}\label{sec:notation} \vspace{-0.15cm}
We indicate column vectors, e.g.~$a$, and matrices, e.g.~$A$, with lowercase and uppercase letters, respectively. We denote stacked column vectors, e.g.~$\mathbf{a}$, and block matrices, e.g.,~$\mathbf{A}$, with bold letters and index their (block) elements with subscripts, e.g.,~$a_i, \ A_{i,j}$. We use~$\diag_i(A)$ to denote the block-diagonal matrix, whose diagonal consists of $i$-times the $A$ matrix. We define the operator $\mathrm{shift}(\mathbf{A},\mathbf{B})$ for a block-lower-triangular matrix~$\mathbf{A}$ with $N \times N$ block elements and a block row matrix~$\mathbf{B}$ with $N$ block elements, i.e.~$\mathbf{B} = [ B_0 \, \mydots \, B_{N-1} ]$, as
\begin{equation*}
        \mathrm{shift}(\mathbf{A},\mathbf{B}) = \begin{bmatrix} A_{1,1} & & & \\ \myvdots & \myddots & & \\ A_{N-1,1} & \mydots & A_{N-1,N-1} & \\ B_0 & \mydots & \mydots & B_{N-1} \end{bmatrix}\!,
\end{equation*}
which removes the first row and column of~$\mathbf{A}$ and inserts~$\mathbf{B}$ in the bottom block row. We use $\mathbb{I}_n$ to denote the identity matrix of size $n \times n$. We distinguish between the states of a dynamical system~$x(k)$ and the states predicted by an MPC algorithm~$x_i$. We use~$\oplus$ and~$\ominus$ to denote Minkowski set addition and Pontryagin set subtraction, respectively, which are both formally defined in~\cite[Definition~3.10]{Rawlings2009}. For a sequence of sets~$\mathcal{A}_0, \mydots, \mathcal{A}_j$, we use the convention that~$\bigoplus_{i=0}^{-1} \mathcal{A}_i = \{0\}$, which is the set containing only the zero vector. For $\mathcal{K}$-functions and uniform continuity we use the standard definitions in~\cite{Rawlings2009}.
\vspace{-0.2cm}
\subsection{Problem Formulation}\vspace{-0.15cm}
We consider linear time-invariant (LTI) dynamical systems with parametric and additive uncertainties, i.e.,
\begin{equation}\label{eq:dynamics}
x(k\!+\!1) = (A + \Delta_A) x(k) + (B + \Delta_B) u(k) + w(k),
\end{equation}
with $A \!\in\! \mathbb{R}^{n \times n}$, $B \!\in\! \mathbb{R}^{n \times m}$, $(\Delta_A, \Delta_B) \!\in\! \mathcal{D}$, and $w(k) \!\in\! \mathcal{W} \subset \mathbb{R}^n$, where $\mathcal{W}$ and $\mathcal{D}$ are the convex polytopes defined as
\begin{subequations}\label{eq:disturbance-sets}
\begin{align}
\mathcal{W} &= \{ w \in \mathbb{R}^n \mid H_{w} w \leq h_{w} \}, \label{eq:W}\\
\mathcal{D} &= \textrm{co} \{(\Delta_A^d, \Delta_B^d)\}, \; d=1,\mydots,n_D,
\end{align}
\end{subequations}
where $H_{w} \in \mathbb{R}^{n_w \times n}$, $h_{w} \in \mathbb{R}^{n_w}$, and $n_D$ is the number of generators for $\Delta_A$ and $\Delta_B$. The set~$\mathcal{D}$ is thus defined by the vertices~$\mathcal{D}^d = (\Delta_A^d,\, \Delta_B^d)$, $d = 1, \mydots, n_D$. The system is subject to compact polytopic state and input constraints
\begin{align}\label{eq:constraints}
\mathcal{X} \!=\! \{ x \!\in\! \mathbb{R}^n \!\mid\! H_{x} x \!\leq\! h_{x} \}, \
\mathcal{U} \!=\! \{ u \!\in\! \mathbb{R}^m \!\mid\! H_{u} u \!\leq\! h_{u} \},
\end{align}
where $H_{x} \in \mathbb{R}^{n_x \times n}$, $h_{x} \in \mathbb{R}^{n_x}$, $H_{u} \in \mathbb{R}^{n_u \times m}$, $h_{u} \in \mathbb{R}^{n_u}$, with both sets containing the origin in their interior.

To safely control system~\eqref{eq:dynamics}, we consider the following robust constraint satisfaction problem (CSP) over the task horizon $\bar{N}$.\vspace{-0.1cm}
\begin{problem}[Robust CSP]\label{prob:CSP}
Find a sequence of inputs~$u_i$, $i=0,\mydots,\bar{N}\!-\!1$, such that the following constraints are satisfied for all $i=0, \mydots, \bar{N}\!-\!1$:
\begin{subequations}\label{eq:CSP}
\begin{align}
        & x_0 = x(k), \label{CSP:init}\\
        & x_{i+1} = (A+\Delta_A)x_i + (B+\Delta_B)u_i + w_i, \label{CSP:prediction-dyn}\\
        & x_i \in \mathcal{X}, \qquad  \forall (\Delta_A, \Delta_B) \!\in \mathcal{D},\ \forall w_i \in \mathcal{W},\\
        & u_i \in \mathcal{U}, \\
        & x_{\bar{N}} \in \mathcal{S}_f,
\end{align}
\end{subequations}
where $x_i$, $u_i$, and $w_i$ denote the predicted state, input, and disturbance, respectively, and $\mathcal{S}_f \subseteq \mathcal{X}$ is a robust positively invariant~(RPI) set according to Definition~\ref{def:RCI}.\!\footnote{Note that we use an RPI set for simplicity, however a robust control invariant (RCI) set could be used instead.}
\end{problem}
\vspace{-0.15cm}
\begin{definition}[RPI set for~\eqref{eq:dynamics}]\label{def:RCI}
The set $\mathcal{S}_f \subseteq \mathcal{X}$ is a robust positively invariant (RPI) set for system~\eqref{eq:dynamics} with control law $u = K_f x \in \mathcal{U}$ for all $x \in \mathcal{S}_f$, if $x \in \mathcal{S}_f \implies x^+ \in \mathcal{S}_f$ for all $w \in \mathcal{W}$, $(\Delta_A, \Delta_B) \in \mathcal{D}$.
\end{definition}\vspace{-0.1cm}
In this paper, we solve the CSP using a tube-based MPC formulation~\cite{Rawlings2009,Sieber2021} that tightens the constraints with so-called tubes for a shorter horizon $N < \bar{N}$ and apply the MPC in shrinking horizon or receding horizon. In the following section, we detail how to derive such a formulation by generalizing the tube-based MPC formulation proposed in~\cite{Chen2023}.
\vspace{-0.17cm}
\section{Filter-based System Level Tube-MPC}\label{sec:SLTMPC}\vspace{-0.12cm}
The filter-based SLTMPC proposed in~\cite{Chen2023} relies on designing an auxiliary disturbance set~$\bar{\mathcal{W}}$, before transforming it online to overapproximate the combined effect of additive and parametric uncertainties. This enables online adaptation of the tube controller and results in a nonconservative tube-based MPC method. We extend this method to general polytopic disturbance sets~$\bar{\mathcal{W}} = \{ \bar{w} \in \mathbb{R}^n \mid H_{\bar{w}} \bar{w} \leq h_{\bar{w}} \}$, with $H_{\bar{w}} \in \mathbb{R}^{n_{\tilde{w}} \times n}, h_{\bar{w}} \in \mathbb{R}^{n_{\tilde{w}}}$, instead of norm-bounded disturbance sets~$\bar{\mathcal{W}} = \{ \bar{w} \in \mathbb{R}^n \mid \| \bar{w} \|_\infty \leq 1 \}$, which extends the scope of the method and simplifies its exposition.
We first rewrite the prediction dynamics~\eqref{CSP:prediction-dyn} by collecting all uncertainties in a separate term denoted as~$\eta_i$:
\begin{align}
x_{i+1} &= A x_i + B u_i + \eta_i, \label{eq:lumped-dyn} \\
\eta_i &= \Delta_A x_i + \Delta_B u_i + w_i. \label{eq:disturbance-dyn}
\end{align}
Following the idea presented in~\cite{Chen2023}, we then overapproximate the combined uncertainty~$\eta_i$ for $i=0,\mydots,{N-1}$ with online-computed \emph{disturbance tubes}, i.e.,
\begin{equation}\label{eq:lumped-unc-containment}
\eta_i \in \{ p_i \} \oplus \mathcal{F}_i\left(\bm{\Sigma}\right) \coloneqq \{ p_i \} \oplus \bigoplus_{j=0}^{i} \Sigma_{i+1,j} \bar{\mathcal{W}},
\end{equation}
where $p_i \in \mathbb{R}^n$, $\Sigma_{i+1,j} \in \mathbb{R}^{n \times n}$ are online optimization variables denoting the nominal disturbance and the disturbance filter, respectively. The compact polytopic disturbance set~$\bar{\mathcal{W}} \subset \mathbb{R}^n$, which contains the origin in its interior, is designed offline. Note that these disturbance tubes~\eqref{eq:lumped-unc-containment} are a generalization of homothetic tubes~\cite{Rakovic2012b}. This can be seen by setting~$\Sigma_{i+1,i} = \alpha_i \cdot \mathbb{I}_n$ and~$\Sigma_{i+1,j} = 0$ for $j=0, \mydots, i-1$, which simplifies containment condition~\eqref{eq:lumped-unc-containment} to $\eta_i \in \{ p_i \} \oplus \alpha_i \bar{\mathcal{W}}$, where~$\alpha_i$ are the dilation factors and $p_i$ are the tube centers in~\cite{Rakovic2012b}. If we enforce~\eqref{eq:lumped-unc-containment} for all~$w \in \mathcal{W}$, $(\Delta_A, \Delta_B) \in \mathcal{D}$ during online optimization, we can guarantee that for optimized~$p_i, \, \Sigma_{i+1,j}$, there always exists a sequence of $\bar{w}_j \in \bar{\mathcal{W}},\, j= 0, \mydots, i$ such that
\begin{equation*}
\eta_i =  p_i + \sum_{j=0}^{i} \Sigma_{i+1,j} \bar{w}_j, \quad i=0,\mydots,N\!-\!1.
\end{equation*}
Therefore, we can equivalently write dynamics~\eqref{eq:lumped-dyn} as
\begin{equation}\label{eq:auxiliary-dynamics}
        x_{i+1} = Ax_i + Bu_i + p_i + \sum_{j=0}^{i} \Sigma_{i+1,j} \bar{w}_j,
\end{equation}
when imposing that~\eqref{eq:lumped-unc-containment} holds.
\begin{remark}
The application of disturbance tubes~\eqref{eq:lumped-unc-containment} is not limited to system \eqref{eq:dynamics} with uncertainty description~$\mathcal{W}, \, \mathcal{D}$, but can be extended to other uncertainty descriptions, e.g., the one used in~\cite{Kohler2019}, as long as the combined uncertainty~$\eta_i$ is a linear function of $x_i$ and $u_i$.
\end{remark}
In the following, we use auxiliary dynamics~\eqref{eq:auxiliary-dynamics} and formulate the filter-based SLTMPC problem by splitting~\eqref{eq:auxiliary-dynamics} into nominal and error dynamics. We use $z_i$ and $e_i$ as the nominal and the error state, respectively, such that $x_i = z_i + e_i$. Together with the tube-based control policy $u_i = v_i + \nu_i$, where $v_i$ and $\nu_i$ are the nominal and error inputs, we obtain
\begin{align*}
        x_{i+1} &= A(z_i + e_i) + B(v_i + \nu_i) + p_i + \sum_{j=0}^{i} \Sigma_{i+1,j} \bar{w}_j,
\end{align*}
and thus -- as is standard in tube-based MPC~\cite{Rawlings2009} -- separation into nominal and error dynamics, i.e.,
\begin{align}
        z_{i+1} &= Az_i + Bv_i + p_i, \label{eq:nominal-dynamics}\\
        e_{i+1} &= Ae_i + B\nu_i + \sum_{j=0}^{i} \Sigma_{i+1,j} \bar{w}_j. \label{eq:error-dynamics}
\end{align}
Since we consider MPC with horizon length~$N$, we define the trajectories $\mathbf{z} = [ z_0^\top, \mydots, z_N^\top ]^\top$, $\mathbf{v} = [ v_0^\top, \mydots, v_N^\top ]^\top$, $\mathbf{p} = [ p_0^\top, \mydots, p_{N-1}^\top ]^\top$, $\bar{\mathbf{w}} = [ \bar{w}_0^\top, \mydots, \bar{w}_{N-1}^\top ]^\top$, $\mathbf{e} = [ e^{\top}_1, \mydots, e^{\top}_{N}]^\top$, and $\bm{\nu} = [ \nu^{\top}_1, \mydots, \nu^{\top}_N ]^\top$ as the nominal state, nominal input, nominal disturbance, auxiliary disturbance, error state,\!\footnote{Note that the initial error state $e_0$ is zero by construction, thus the trajectory $\mathbf{e}$ starts with $e_1$.} and error input trajectories, respectively. The nominal and error dynamics are then compactly written in terms of these trajectories as
\begin{align}
\mathbf{z} &= \mathbf{ZA}_{N+1}\mathbf{z} + \mathbf{ZB}_{N+1}\mathbf{v} + \begin{bmatrix} z_0 \\ \mathbf{p}\end{bmatrix}, \\
\mathbf{e} &= \mathbf{ZA}_{N}\mathbf{e} + \mathbf{ZB}_{N}\bm{\nu} + \bm{\Sigma}\bar{\mathbf{w}}, \label{eq:stacked-error-dynamics}
\end{align}
where $\mathbf{Z}$ is a lower shift matrix, i.e., a matrix with ones on its first subdiagonal and zeros everywhere else, $\mathbf{A}_{N} = \diag_{N}(A)$, $\mathbf{B}_{N} = \diag_{N}(B)$, and~$\bm{\Sigma}$ is the block-lower-triangular \emph{disturbance filter} defined as
\begin{equation*}
\bm{\Sigma} \!=\! \begin{bmatrix} \Sigma_{1,0} & & & \\ \Sigma_{2,0} & \Sigma_{2,1} &  &  \\ \myvdots & \myvdots & \myddots & \\ \Sigma_{N,0} & \Sigma_{N,1} & \mydots & \Sigma_{N,N-1} \end{bmatrix} \in \mathbb{R}^{Nn\times Nn}.
\end{equation*}
Next, we define the error input as the causal time-varying feedback controller $\bm{\nu} = \mathbf{K}\mathbf{e}$ with
\begin{equation*}
\mathbf{K} = \begin{bmatrix} K_{1,1} & & & \\ K_{2,1} & K_{2,2} &  &  \\ \myvdots & \myvdots & \myddots & \\ K_{N,1} & K_{N,2} & \mydots & K_{N,N} \end{bmatrix} \in \mathbb{R}^{Nm\times Nn}.
\end{equation*}
Substituting this error input into~\eqref{eq:stacked-error-dynamics} yields
\begin{align}
&\mathbf{e}= \left(\mathbf{ZA}_{N} + \mathbf{ZB}_{N}\mathbf{K}\right)\mathbf{e} + \bm{\Sigma} \bar{\mathbf{w}}, \nonumber \\
\Leftrightarrow\ &\mathbf{e} = \left( \mathbb{I}_{Nn} - \mathbf{ZA}_{N} - \mathbf{ZB}_{N}\mathbf{K}\right)^{-1}\bm{\Sigma} \bar{\mathbf{w}}. \label{eq:SLP-error-dyn}
\end{align}
Note that the inverse in~\eqref{eq:SLP-error-dyn} always exists, since it admits a finite Neumann series due to the lower shift matrix~$\mathbf{Z}$. As in~\cite{Chen2023}, we use the filter-based system level parameterization~(SLP) to reparameterize~\eqref{eq:SLP-error-dyn} and the associated error input~$\bm{\nu}$ as
\begin{subequations}\label{eq:SLP}
\begin{align}
\mathbf{e} &= \left( \mathbb{I}_{Nn} - \mathbf{ZA}_{N} - \mathbf{ZB}_{N}\mathbf{K}\right)^{-1}\bm{\Sigma}\bar{\mathbf{w}} = \bm{\Phi}^\mathbf{e}\,\bar{\mathbf{w}}, \\
\bm{\nu} &= \mathbf{K}\left( \mathbb{I}_{Nn} - \mathbf{ZA}_{N} - \mathbf{ZB}_{N}\mathbf{K}\right)^{-1}\bm{\Sigma}\bar{\mathbf{w}} = \bm{\Phi}^{\bm{\nu}}\bar{\mathbf{w}},
\end{align}
\end{subequations}
where $\bm{\Phi}^\mathbf{e} \in \mathbb{R}^{Nn\times Nn}, \bm{\Phi}^{\bm{\nu}} \in \mathbb{R}^{Nm\times Nn}$ are the \emph{error system responses}, which admit the following block-lower-triangular structure\footnote{Note that the indexing of $\Phi^{e}_{i,j}$ is different from the usual indexing in~\cite{Sieber2021},~\cite{Anderson2019}, or \cite{Chen2023} in order to reflect that $\Phi^{e}_{i,j}$ affects the $i$-th error state through the $j$-th auxiliary disturbance.}\vspace{-0.15cm}
\begin{align*}
\bm{\Phi}^\mathbf{e} \!=\!\! \begin{bmatrix} \!\Phi^{e}_{1,0} & & & \\ \!\Phi^{e}_{2,0} & \!\Phi^{e}_{2,1} &  &  \\ \!\myvdots & \!\myvdots & \!\myddots & \\ \!\Phi^{e}_{\!N,0} & \!\Phi^{e}_{\!N,1} & \!\mydots & \!\Phi^{e}_{\!N,N\textrm{-}1} \end{bmatrix}\!\!, \,
\bm{\Phi}^{\bm{\nu}} \!\!=\!\! \begin{bmatrix} \!\Phi^{\nu}_{1,0} & & & \\ \!\Phi^{\nu}_{2,0} & \!\Phi^{\nu}_{2,1} &  &  \\ \!\myvdots & \!\myvdots & \!\myddots & \\ \!\Phi^{\nu}_{\!N,0} & \!\Phi^{\nu}_{\!N,1} & \!\mydots & \!\Phi^{\nu}_{\!N,N\textrm{-}1} \end{bmatrix}\!\!.
\vspace{-0.15cm}
\end{align*}
Using~\cite[Theorem~2.1]{Anderson2019} and~\cite[Corollary~1]{Chen2023} we guarantee that the error system responses $\bm{\Phi}^\mathbf{e}, \bm{\Phi}^{\bm{\nu}}$ parameterize all error trajectories $\mathbf{e}, \bm{\nu}$ that are realized by tube controller $\mathbf{K}$ and disturbance filter~$\bm{\Sigma}$. Using the filter-based SLP~\eqref{eq:SLP}, the error dynamics~\eqref{eq:stacked-error-dynamics} are equivalently rewritten as\vspace{-0.15cm}
\begin{equation}\label{eq:SLP-affine-constraint}
\begin{bmatrix} \mathbb{I}_{Nn} \!-\! \mathbf{ZA}_N & \ \scalebox{0.85}[1.0]{$-$}\mathbf{ZB}_N \end{bmatrix} \!\begin{bmatrix} \bm{\Phi}^\mathbf{e} \\ \bm{\Phi}^{\bm{\nu}} \end{bmatrix} = \bm{\Sigma}.
\vspace{-0.15cm}
\end{equation}
Thereby, the error dynamics are completely defined by the closed-loop behavior, which allows optimization over the system responses in a convex fashion instead of optimizing the tube controller. Note that the filter-based SLP~\eqref{eq:SLP} can also be written in elementwise form, i.e., for~$i=1,\mydots,N$\vspace{-0.15cm}
\begin{equation}\label{eq:stagewise-SLP}
        e_i = \sum_{j=0}^{i-1} \Phi^{e}_{i,j}\bar{w}_j, \qquad
        \nu_i = \sum_{j=0}^{i-1} \Phi^{\nu}_{i,j}\bar{w}_j,
        \vspace{-0.15cm}
\end{equation}
and that~\eqref{eq:SLP-affine-constraint} constrains the block diagonals of $\bm{\Phi}^\mathbf{e}$ and $\bm{\Sigma}$ due to the lower shift matrices~$\mathbf{Z}$, i.e., it holds that $\Phi^{e}_{i+1,i} = \Sigma_{i+1,i}$ for all $i = 0,\mydots,N\!-\!1$.
In~\cite[Section~4]{Chen2023} it is shown how to enforce containment condition~\eqref{eq:lumped-unc-containment} for norm-bounded auxiliary disturbance sets~$\bar{\mathcal{W}}$. We extend this analysis to general polytopic disturbance sets~$\bar{\mathcal{W}}$ in the following. First, we split the state~$x_i$ and input~$u_i$ in~\eqref{eq:disturbance-dyn} into their nominal and error contributions and apply the elementwise SLP~\eqref{eq:stagewise-SLP} to the error state~$e_i$ and error input $\nu_i$, i.e.,
\begin{align*}
\eta_i &= \Delta_A (z_i + e_i) + \Delta_B (v_i + \nu_i) + w_i \\
&= \Delta_A z_i + \Delta_B v_i + \sum_{j=0}^{i-1} \left( \Delta_A \Phi^{e}_{i,j} + \Delta_B \Phi^{\nu}_{i,j} \right)\bar{w}_j + w_i.
\end{align*}
The containment condition~\eqref{eq:lumped-unc-containment} for $i=0,\mydots, N\!-\!1$ then reads as
\begin{align*}
&\{ \Delta_A z_i + \Delta_B v_i \} \oplus \bigoplus_{j=0}^{i-1}\left(\Delta_A \Phi^{e}_{i,j} + \Delta_B \Phi^{\nu}_{i,j} \right) \bar{\mathcal{W}} \oplus \mathcal{W} \\
&\subseteq \{ p_i \} \oplus \bigoplus_{j=0}^{i}\Sigma_{i+1,j} \bar{\mathcal{W}}, \quad \forall (\Delta_A , \Delta_B) \in \mathcal{D},
\end{align*}
and by collecting the singletons we get \vspace{-0.1cm}
\begin{align}\label{eq:uncertainty-inclusion}
&\{ \Delta_A z_i + \Delta_B v_i - p_i \} \oplus \bigoplus_{j=0}^{i-1} (\Delta_A \Phi^{e}_{i,j} + \Delta_B \Phi^{\nu}_{i,j}) \bar{\mathcal{W}} \oplus \mathcal{W} \nonumber\\
&\subseteq \bigoplus_{j=0}^{i} \Sigma_{i+1,j}\bar{\mathcal{W}} = \mathcal{F}_i(\bm{\Sigma}), \quad \forall (\Delta_A , \Delta_B) \in \mathcal{D}.
\vspace{-0.1cm}
\end{align}
Since the superset of above inclusion is a Minkowski sum, embedding it in an optimization problem is conservative and nonconvex~\cite{Sadraddini2019}. Therefore, we make use of the following lemma to convert~\eqref{eq:uncertainty-inclusion} into a form that can be formulated linearly in the optimization variables. \vspace{-0.1cm}
\begin{lem}\label{lem:inclusion}
Let~$\mathcal{D}^d = (\Delta_A^d, \Delta_B^d)$ denote the $d^\textrm{th}$ vertex of~$\mathcal{D}$ and define \vspace{-0.1cm}
\begin{subequations}\label{eq:q_Pi}
\begin{align}
\psi^d_i &\coloneqq \Delta_A^d z_i + \Delta_B^d v_i -p_i, \label{eq:q}\\
\Psi_{i,j}^d &\coloneqq \Delta_A^d \Phi^{e}_{i,j} + \Delta_B^d \Phi^{\nu}_{i,j} - \Sigma_{i+1,j}. \label{eq:Pi}
\vspace{-0.1cm}
\end{align}
\end{subequations}
Then, if the following inclusion holds for $i=0,\mydots,N\!-\!1$ and $d=1,\mydots,n_D$,\vspace{-0.1cm}
\begin{equation}\label{lemma:inclusion}
\{ \psi^d_i \} \oplus \bigoplus_{j=0}^{i-1}\Psi^d_{i,j}\bar{\mathcal{W}} \oplus \mathcal{W} \subseteq \Sigma_{i+1,i}\,\bar{\mathcal{W}},
\vspace{-0.1cm}
\end{equation}
containment condition~\eqref{eq:uncertainty-inclusion} is satisfied.
\end{lem}
\vspace{-0.2cm}
\begin{pf}
First, note that due to the separation $x = z + e$ and $u = v + \nu$, the same $\Delta_A^d$ and $\Delta_B^d$ are applied to the nominal and error variables. Therefore, the same vertex of~$\mathcal{D}$ is used in~\eqref{eq:q},~\eqref{eq:Pi} and we can prove the lemma for any vertex~$d$, implying that the proof holds for all~$(\Delta_A, \, \Delta_B) \in \mathcal{D}$ due to convexity of~$\mathcal{D}$. We start by rewriting~\eqref{eq:uncertainty-inclusion} using~\eqref{eq:q_Pi} as
\begin{equation}\label{proof:rewritten-inclusion}
\{ \psi^d_i \} \oplus \bigoplus_{j=0}^{i-1} (\Psi^d_{i,j} + \Sigma_{i+1,j}) \bar{\mathcal{W}} \oplus \mathcal{W} \subseteq \bigoplus_{j=0}^{i} \Sigma_{i+1,j}\bar{\mathcal{W}},
\end{equation}
where we used $\Psi^d_{i,j} + \Sigma_{i+1,j} = \Delta^d_A \Phi^{e}_{i,j} + \Delta^d_B \Phi^{\nu}_{i,j}$. Next, we note that the following inclusion holds
\begin{equation}\label{proof:helping-lemma}
\left( \Psi^d_{i,j} + \Sigma_{i+1,j}\right)\bar{\mathcal{W}} \subseteq  \Psi^d_{i,j}\bar{\mathcal{W}} \oplus \Sigma_{i+1,j}\bar{\mathcal{W}},
\end{equation}
which is a direct consequence of~\cite[Proposition~2]{Sadraddini2019}. Then, we show~\eqref{proof:rewritten-inclusion} holds by using~\eqref{proof:helping-lemma} and commutativity of the Minkowski sum, i.e.,
\begin{align*}
&\{ \psi^d_i \} \oplus \bigoplus_{j=0}^{i-1} (\Psi^d_{i,j} + \Sigma_{i+1,j}) \bar{\mathcal{W}} \oplus \mathcal{W} \\
&\overset{\mathclap{\strut\text{\eqref{proof:helping-lemma}}}}\subseteq \{ \psi^d_i \} \oplus \bigoplus_{j=0}^{i-1} \Psi^d_{i,j}\bar{\mathcal{W}} \oplus \mathcal{W} \oplus \bigoplus_{j=0}^{i-1} \Sigma_{i+1,j} \bar{\mathcal{W}} \\
&\overset{\mathclap{\strut\text{\eqref{lemma:inclusion}}}}\subseteq \Sigma_{i+1,i}\bar{\mathcal{W}} \oplus \bigoplus_{j=0}^{i-1} \Sigma_{i+1,j} \bar{\mathcal{W}}.
\end{align*}
Finally, we note that the right-hand side of above inclusion can be rewritten as a single Minkowski sum, i.e.,
\begin{align*}
        \Sigma_{i+1,i}\bar{\mathcal{W}} \oplus \bigoplus_{j=0}^{i-1} \Sigma_{i+1,j} \bar{\mathcal{W}} = \bigoplus_{j=0}^{i} \Sigma_{i+1,j} \bar{\mathcal{W}}
\end{align*}
and thus we have shown that enforcing~\eqref{lemma:inclusion} is sufficient for~\eqref{eq:uncertainty-inclusion} to hold.\qed
\end{pf}
\vspace{-0.15cm}
Additionally, we impose a diagonal structure on $\bm{\Sigma}$, i.e., $\Sigma_{i+1,i} = \sigma_i \cdot \mathbb{I}_{n}$ for $i=0,\mydots,N\!-\!1$, which is similarly done in~\cite[Section~4.1]{Chen2023}. Using the correspondence to homothetic tubes~\cite{Rakovic2012b}, we note that this diagonal structure with scalings~$\sigma_i$ is equivalent to the dilation factors~$\alpha_i$. After applying Lemma~\ref{lem:inclusion} and imposing the diagonal structure, the inclusion~\eqref{eq:uncertainty-inclusion} for $i=0,\mydots, N\!-\!1$ and $d=1,\mydots,n_D$ reads as
\begin{align}\label{eq:linear-uncertainty-inclusion}
\{ \psi^d_i \} \oplus \bigoplus_{j=0}^{i-1} \Psi^d_{i,j} \bar{\mathcal{W}} \oplus \mathcal{W} \subseteq \sigma_{i+1}\bar{\mathcal{W}},
\end{align}
which can be linearly embedded in an optimization problem as shown in Appendix~\ref{app:implementation}.
\begin{remark}\label{remark:hyperrectangle}
Note that we can relax the structural constraint~$\Sigma_{i+1,i} = \sigma_i \cdot \mathbb{I}_n$ to $\Sigma_{i+1,i} = \textrm{diag}\,(\sigma_{i,1}, \mydots, \sigma_{i,n})$, if we design~$\bar{\mathcal{W}}$ as a hyperrectangle, therefore recovering the same disturbance tube parameterization as in~\cite{Chen2023}. For more details, see Appendix~\ref{app:implementation}.
\end{remark}
Finally, we use nominal dynamics~\eqref{eq:nominal-dynamics}, error dynamics~\eqref{eq:SLP-affine-constraint}, and disturbance overapproximation~\eqref{eq:linear-uncertainty-inclusion} to formulate the generalized version of the filter-based SLTMPC problem~\cite{Chen2023} as
\begin{subequations}\label{SLTMPC:generic}
        \begin{align}
                \min_{\substack{\mathbf{z}, \mathbf{v}, \mathbf{p}, \bm{\Sigma},\\\bm{\Phi}^\mathbf{e}, \bm{\Phi}^{\bm{\nu}}}} \quad & l_f(z_N) + \sum_{i=0}^{N-1} l(z_i, v_i), \label{SLTMPC:cost}\\
                \textrm{s.t. } \:\; & \forall\, i=0, \mydots, N\!-\!1\!: \nonumber\\
                & z_0 = x(k), \label{SLTMPC:init}\\
                & z_{i+1} = Az_i + Bv_i + p_i, \\
                & \begin{bmatrix} \mathbb{I}_{Nn} \!-\! \mathbf{ZA}_N & \ \scalebox{0.85}[1.0]{$-$}\mathbf{ZB}_N \end{bmatrix} \!\begin{bmatrix} \bm{\Phi}^\mathbf{e} \\ \bm{\Phi}^{\bm{\nu}} \end{bmatrix} = \bm{\Sigma}, \\[0.12cm]
                & z_i \in \mathcal{X} \ominus \mathcal{F}_i\left( \bm{\Phi}^\mathbf{e} \right), \\
                & v_i \in \mathcal{U} \ominus \mathcal{F}_i\left( \bm{\Phi}^{\bm{\nu}} \right), \\
                & z_N \in \mathcal{S}_f \ominus \mathcal{F}_N\left( \bm{\Phi}^\mathbf{e} \right), \\[0.12cm]
                & \{ \psi^d_i \} \!\oplus\! \bigoplus_{j=0}^{i-1} \Psi^d_{i,j} \bar{\mathcal{W}} \!\oplus\! \mathcal{W} \!\subseteq\! \sigma_{i+1}\bar{\mathcal{W}}, \  d = 1, \mydots, n_D,
        \end{align}
\end{subequations}
where $l(\cdot, \cdot)$ and $l_f(\cdot)$ are suitable stage and terminal costs, $\psi^d_i$ and $\Psi^d_{i,j}$ are defined as in~\eqref{eq:q_Pi}, $\mathcal{S}_f$ is an RPI terminal set according to Definition~\ref{def:RCI}, and $\mathcal{F}_i\left( \bm{\Phi}^\mathbf{e} \right)$, $\mathcal{F}_i\left( \bm{\Phi}^{\bm{\nu}} \right)$ are the state and input tubes defined as \vspace{-0.15cm}
\begin{align}\label{eq:tubes}
\mathcal{F}_i\left(\bm{\Phi}^\mathbf{e}\right) &\coloneqq \bigoplus_{j=0}^{i-1} \Phi^{e}_{i,j} \bar{\mathcal{W}}, \quad \mathcal{F}_i\left(\bm{\Phi}^{\bm{\nu}}\right) \coloneqq \bigoplus_{j=0}^{i-1} \Phi^{\nu}_{i,j} \bar{\mathcal{W}}.
\vspace{-0.15cm}
\end{align}
In order to show recursive feasibility of~\eqref{SLTMPC:generic} we would need to show that $\mathcal{S}_f \ominus \mathcal{F}_N(\bm{\Phi}^\mathbf{e})$ is RPI itself. However, since $\mathcal{S}_f$ is RPI with respect to both $w \in \mathcal{W}$ and $(\Delta_A, \Delta_B) \in \mathcal{D}$ (Definition~\ref{def:RCI}) this is difficult because $\mathcal{F}_N(\bm{\Phi}^\mathbf{e})$ would need to be an exact reachable set of~\eqref{eq:dynamics} for all $w \in \mathcal{W},\, (\Delta_A, \Delta_B) \in \mathcal{D}$. This is clearly not the case, since $\mathcal{F}_N(\bm{\Phi}^\mathbf{e})$ is computed via~\eqref{eq:uncertainty-inclusion}, which is an overapproximation of the combined uncertainties. Therefore, we need to restrict the SLTMPC~\eqref{SLTMPC:generic} to a shrinking horizon regime similar to~\cite{Bujarbaruah2022,Chen2023}, which switch between solving the MPC problem with a shrinking horizon and exactly solving the robust CSP~\eqref{eq:CSP} for $\bar{N}=1$ to show recursive feasibility and robust stability. However, this strategy requires implementation of a switching logic and only works well if task horizon $\bar{N}$ is finite and known in advance. In the next section, we show how to modify the terminal constraints in~\eqref{SLTMPC:generic} such that the resulting MPC can be applied in receding horizon. For a recursive feasibility proof of~\eqref{SLTMPC:generic} in shrinking horizon, we refer to~\cite[Appendix~6]{Bujarbaruah2022}.
\vspace{-0.2cm}
\section{Recursively Feasible Filter-based SLTMPC}\label{sec:SLTMPC-rec-feas}\vspace{-0.18cm}
The terminal constraints in~\eqref{SLTMPC:generic} are not suitable to prove recursive feasibility in receding horizon, due to~$\mathcal{S}_f$ being computed for system~\eqref{eq:dynamics} with combined uncertainty~$\eta$, and $\mathcal{F}_N(\bm{\Phi}^\mathbf{e})$ being computed for auxiliary system~\eqref{eq:auxiliary-dynamics} with only additive uncertainty~$\bar{w}$. Therefore, we propose a new set of terminal constraints that only rely on sets computed for auxiliary system~\eqref{eq:auxiliary-dynamics}. The key idea is to exploit the separation~$x = z + e$ and formulate a separate terminal control law for both the nominal and error states, ensuring that both only depend on~$\bar{w}$. The resulting new terminal set~$\mathcal{X}_f$ is then used to constrain the terminal state~$x_N \in \mathcal{X}_f$ in the proposed MPC scheme.
For this we first define an auxiliary RPI set~$\mathcal{Z}_f$ for a simplified version of~\eqref{eq:auxiliary-dynamics} with $p_i=0$, $\Sigma_{i+1,i} = \mathbb{I}_n, \, \Sigma_{i+1, j} = 0$ for all $i$ and $j=0, \mydots, i\!-\!1$, i.e.,\vspace{-0.05cm}
\begin{equation}\label{eq:simplified-auxiliary-dynamics}
        x(k+1) = Ax(k) + Bu(k) + \bar{w}(k).
\end{equation}
\vspace{-0.6cm}
\begin{definition}[RPI set for~\eqref{eq:simplified-auxiliary-dynamics}]\label{def:RPI}
The set $\mathcal{Z}_f \subseteq \mathcal{X}$ is a robust positively invariant (RPI) set for system~\eqref{eq:simplified-auxiliary-dynamics} with control law $u = K_f x \in \mathcal{U}$ for all~$x \in \mathcal{Z}_f$, if $x \in \mathcal{Z}_f \implies x^+ \in \mathcal{Z}_f$ for all $\bar{w}(k) \in \bar{\mathcal{W}}$.
\end{definition}
\vspace{-0.1cm}
We then use the control law~$K_f$ of RPI set~$\mathcal{Z}_f$ to construct the terminal control law as\vspace{-0.1cm}
\begin{align}\label{eq:terminal-control-law}
        \kappa_f(x) \!=\! \kappa_f^z(z) \!+\! \kappa_f^e(e) \!=\! K_f z \!+\! \nu \!=\! K_f z \!+\!\! \sum_{j = 0}^{N} \Phi^{\nu}_{N,j} \bar{w}_{N-j},
        \vspace{-0.1cm}
\end{align}
where $\Phi^{\nu}_{N,j}$, i.e. the last block row of $\bm{\Phi}^{\bm{\nu}}$, can be freely optimized since it is not constrained by~\eqref{eq:SLP-affine-constraint}. Before constructing the RPI set corresponding to control law~\eqref{eq:terminal-control-law}, we first overapproximate the terminal uncertainty~$\eta_N$ with a terminal disturbance tube similar to~\eqref{eq:lumped-unc-containment} -- since $i=N$ is not included in~\eqref{eq:lumped-unc-containment} -- i.e.,\vspace{-0.1cm}
\begin{align}\label{eq:terminal-dist-tube}
\eta_N \in \mathcal{F}_N\left(\bm{\Xi}\right) \coloneqq \bigoplus_{j=0}^{N-1} \Xi_{j} \bar{\mathcal{W}} \oplus \sigma_N\bar{\mathcal{W}},
\vspace{-0.1cm}
\end{align}
where $\bm{\Xi} = [ \Xi_0 \ \mydots \ \Xi_{N-1} ]$ are additional disturbance filter parameters. Similar to Section~\ref{sec:SLTMPC}, there always exists a sequence of $\bar{w}_j \in \bar{\mathcal{W}}, \, j=0, \mydots N$, such that $\eta_N = \sum_{j=0}^{N-1}\Xi_j \bar{w}_j + \sigma_N\bar{w}_N$. With this, we define RPI set~$\mathcal{X}_f$ for control law~\eqref{eq:terminal-control-law} in the following lemma. \vspace{-0.1cm}
\begin{lem}\label{lem:terminal-RPI-set}
The set $\mathcal{X}_f \coloneqq \alpha \mathcal{Z}_f \oplus \mathcal{F}_N(\bm{\Phi}^\mathbf{e})$, with $\alpha \geq 0$, is an RPI set for system~$x_{N+1} = Ax_N + B\kappa_f(x_N) + \eta_N$ with control law~$\kappa_f(\cdot)$ defined in~\eqref{eq:terminal-control-law} and $\eta_N \in \mathcal{F}_N\left(\bm{\Xi}\right)$, if the following two conditions hold: \vspace{-0.1cm}
\begin{enumerate}
        \item[(a)] $\alpha (A + BK_f)\mathcal{Z}_f \,\oplus\, \Gamma \bar{\mathcal{W}} \subseteq \alpha \mathcal{Z}_f$ with $\mathcal{Z}_f$ RPI according to Definition~\ref{def:RPI} and $\Gamma = A \Phi^{e}_{N,0} + B \Phi^{\nu}_{N,0} + \Xi_0$,
        \item[(b)] $\Phi^{e}_{N,j-1} = A\Phi^{e}_{N,j} + B\Phi^{\nu}_{N,j} + \Xi_j, \ \forall j=1,\mydots,N\!-\!1$.
\end{enumerate}
\vspace{-0.1cm}
It holds that if $x_N \in \mathcal{X}_f \implies x_{N+1} \in \mathcal{X}_f\ \forall\;\! \eta_N \in \mathcal{F}_N(\bm{\Xi})$.
\end{lem}
\vspace{-0.3cm}
\begin{pf}
We start by analyzing the dynamics under control law~\eqref{eq:terminal-control-law}, i.e.,\vspace{-0.1cm}
\begin{align*}
        x_{N+1} &= (A+BK_f)z_N + Ae_N + B\nu_N + \eta_N.
        \vspace{-0.1cm}
\end{align*}
Next, we insert the definitions of $e_N, \,\nu_N$, and $\eta_N$ to get\vspace{-0.1cm}
\begin{align*}
        Ae_N \!+\! B\nu_N \!+\! \eta_N &= \!\!\sum_{j=0}^{N-1} ( A \Phi^{e}_{N,j} \!+\! B \Phi^{\nu}_{N,j} \!+\! \Xi_j) \bar{w}_{j} \!+\! \sigma_N \bar{w}_{N} \\
        &= (A \Phi^{e}_{N,0} + B \Phi^{\nu}_{N,0} + \Xi_0) \bar{w}_{0} \\
        &\, + \!\!\sum_{j=1}^{N-1}\!( A \Phi^{e}_{N,j} \!+\! B \Phi^{\nu}_{N,j} \!+\! \Xi_j) \bar{w}_{j} \!+\! \sigma_N \bar{w}_{N}.
        \vspace{-0.1cm}
\end{align*}
Using $\Phi^{e}_{N,N-1} = \Sigma_{N,N-1} = \sigma_N \cdot \mathbb{I}_{n}$ due to~\eqref{eq:SLP-affine-constraint}, $\Gamma = A \Phi^{e}_{N,0} + B \Phi^{\nu}_{N,0} + \Xi_0$ from (a), and (b) we obtain\vspace{-0.1cm}
\begin{align*}
        Ae_N \!+\! B\nu_N \!+\! \eta_N &= \Gamma \bar{w}_{0} + \sum_{j=1}^{N-1} \Phi^{e}_{N,j-1} \bar{w}_{j} + \Phi^{e}_{N,N-1} \bar{w}_{N} \\
        &= \Gamma \bar{w}_{0} + \sum_{j=0}^{N-1} \Phi^{e}_{N,j} \bar{w}_{j+1}.
        \vspace{-0.1cm}
\end{align*}
Therefore, the dynamics under control law~\eqref{eq:terminal-control-law} are\vspace{-0.1cm}
\begin{align*}
        x_{N+1} &= \underbrace{(A +BK_f)z_N + \Gamma \bar{w}_{0}}_{\coloneqq z_{N+1}} + \underbrace{\sum_{j=0}^{N-1} \Phi^{e}_{N,j} \bar{w}_{j+1}}_{\coloneqq e_{N+1}}.
        \vspace{-0.1cm}
\end{align*}
Now, if $x_N = z_N + e_N \in \alpha \mathcal{Z}_f \oplus \mathcal{F}_N(\bm{\Phi}^\mathbf{e}) = \mathcal{X}_f$, then $z_{N+1} \in \alpha \mathcal{Z}_f$ is guaranteed by (a) and $e_{N+1} \in \mathcal{F}_N(\bm{\Phi}^\mathbf{e})$ holds by construction. Therefore, $x_{N+1} \in \mathcal{X}_f$.\qed
\end{pf}
\vspace{-0.1cm}
Intuitively,~(a) in Lemma~\ref{lem:terminal-RPI-set} ensures that $\alpha \mathcal{Z}_f$ is RPI with respect to control law~$K_f$ and optimized disturbance~$\Gamma\bar{w}$, while~(b) constrains the error states to the same tube~$\mathcal{F}_N(\bm{\Phi}^\mathbf{e})$ after $N$ timesteps, which is less restrictive but similar to the finite impulse response (FIR) constraint in~\cite{Sieber2022}. To use~$\mathcal{X}_f$ as the terminal set in the proposed MPC,~$\mathcal{X}_f$ needs to satisfy the state constraints, i.e.~$\mathcal{X}_f \subseteq \mathcal{X}$, and control law~\eqref{eq:terminal-control-law} needs to satisfy the input constraints, i.e.~$\kappa_f(x) \in \mathcal{U}$, for all~$x \in \mathcal{X}_f$. This can be enforced by \vspace{-0.1cm}
\begin{align}
\alpha \mathcal{Z}_f &\subseteq \mathcal{X} \ominus \mathcal{F}_N\left( \bm{\Phi}^\mathbf{e} \right), \label{tsc:state-constraints}\\
\alpha K_f\mathcal{Z}_f &\subseteq \mathcal{U} \ominus \mathcal{F}_N\left( \bm{\Phi}^{\bm{\nu}} \right), \label{tsc:input-constraints}
\vspace{-0.1cm}
\end{align}
where~\eqref{tsc:state-constraints} follows from $\mathcal{X}_f = \alpha \mathcal{Z}_f \oplus \mathcal{F}_N(\bm{\Phi}^\mathbf{e})$ and~\eqref{tsc:input-constraints} follows from \vspace{-0.1cm}
\begin{align*}
\kappa_f(x_N) &= K_f \underbrace{z_N}_{\in \alpha \mathcal{Z}_f} + \underbrace{\sum_{j = 0}^{N} \Phi^{\nu}_{N,j} \bar{w}_{N-j}}_{\in \mathcal{F}_N( \bm{\Phi}^{\bm{\nu}})} \in \mathcal{U}.
\vspace{-0.1cm}
\end{align*}
Note that constraints\eqref{tsc:state-constraints}, \eqref{tsc:input-constraints}, and~(a) in Lemma~\ref{lem:terminal-RPI-set} are the same constraints as those used in~\cite{Sieber2023} to scale the terminal set. Hence, we can use the results in~\cite{Sieber2023} to linearly embed these constraints into the MPC formulation.

Finally, we reformulate the inclusion of the terminal disturbance~\eqref{eq:terminal-dist-tube} similar to Section~\ref{sec:SLTMPC}. We enforce inclusion~\eqref{eq:terminal-dist-tube} on the combined disturbance~\eqref{eq:disturbance-dyn} with
\begin{align}
&\{ (\underbrace{\Delta_A^d \!+\! \Delta_B^d K_f}_{\coloneqq \Delta^d_{K_f}})z_N \} \oplus \bigoplus_{j=0}^{N-1}\!\left(\Delta_A^d \Phi^{e}_{N,j} \!+\! \Delta_B^d \Phi^{\nu}_{N,j} \right)\! \bar{\mathcal{W}} \oplus \mathcal{W} \nonumber\\
&\subseteq \!\bigoplus_{j=0}^{N-1}\Xi_{j} \bar{\mathcal{W}} \oplus \sigma_N\bar{\mathcal{W}} = \mathcal{F}_N\left(\bm{\Xi}\right), \quad d=1, \mydots, n_D. \label{eq:terminal-dist-inclusion}
\end{align}
Using the terminal constraint~$z_N \in \alpha \mathcal{Z}_f$ and Lemma~\ref{lem:inclusion}, we get
\begin{align*}
&\alpha \Delta^d_{K_f} \mathcal{Z}_f \oplus \bigoplus_{j=0}^{N-1}\left(\Delta^d_A \Phi^{e}_{N,j} + \Delta^d_B \Phi^{\nu}_{N,j} - \Xi_{j} \right) \bar{\mathcal{W}} \oplus \mathcal{W} \\
&\subseteq \sigma_{N}\bar{\mathcal{W}}, \quad d=1, \mydots, n_D,
\end{align*}
or in compact notation using~$\Psi^d_{N,j} = \Delta_A^d \Phi^{e}_{N,j} + \Delta_B^d \Phi^{\nu}_{N,j} - \Xi_{j}$:
\begin{equation}\label{eq:linear-terminal-dist-inclusion}
\alpha \Delta^d_{K_f}\mathcal{Z}_f \oplus \bigoplus_{j=0}^{N-1} \!\Psi^d_{N,j} \bar{\mathcal{W}} \oplus \mathcal{W} \!\subseteq \sigma_{N}\bar{\mathcal{W}}, \ \; d=1,\mydots, n_D. 
\end{equation}

\begin{remark}\label{remark:SLTMPC-hyperrectangle}
Since we derived terminal disturbance inclusion~\eqref{eq:linear-terminal-dist-inclusion} similar to~\eqref{eq:linear-uncertainty-inclusion} in Section~\ref{sec:SLTMPC}, Remark~\ref{remark:hyperrectangle} also applies to~\eqref{eq:linear-terminal-dist-inclusion} and thus allows to replace $\sigma_N$ with $\diag(\sigma_{N,1}, \mydots, \sigma_{N,n})$ if $\bar{\mathcal{W}}$ is designed as a hyperrectangle. Such a choice for~$\bar{\mathcal{W}}$ thus increases the flexibility of the SLTMPC problem proposed below. However, depending on the disturbance sets~$\mathcal{W}$, $\mathcal{D}$, choosing $\bar{\mathcal{W}}$ as a hyperrectangle might be conservative.
\end{remark}
Before formulating the proposed filter-based SLTMPC problem, we state some standard assumptions on the nominal cost function.
\begin{assumption}\label{assump:cost}
The stage cost function $l(x,u)$ is convex, satisfies~$l(0,0) =0$, is uniformly continuous for all $x \in \mathcal{X}, u \in \mathcal{U}$, and there exists a $\mathcal{K}$-function $\gamma_1(\cdot)$ satisfying $l(x,u) \geq \gamma_1(\|x\|), \forall x \in \mathcal{X}, u \in \mathcal{U}$. The terminal cost $l_f(x)$ is convex, positive definite, uniformly continuous for all $x \in \mathcal{X}_f$, there exists a $\mathcal{K}$-function $\gamma_2(\cdot)$ satisfying $l_f(x) \leq \gamma_2(\|x\|), \forall x \in \mathcal{X}_f$, and it holds that $l_f(x^+) - l_f(x) \leq -l(x, K_f x)$ with $x^+ = Ax + B K_fx, \forall x \in \mathcal{X}_f$ and $K_f x \in \mathcal{U}$.
\end{assumption}

The recursively feasible version of the filter-based SLTMPC problem is then formulated as
\begin{subequations}\label{SLTMPC:rec-feas}
    \begin{align}
        \min_{\substack{\mathbf{z}, \mathbf{v}, \mathbf{p}, \alpha \geq 0, \\ \bm{\Phi}^\mathbf{e}, \bm{\Phi}^{\bm{\nu}}, \bm{\Sigma}, \bm{\Xi}}} \quad & l_f(z_N) + \sum_{i=0}^{N-1} l(z_i, v_i), \label{SLTMPC-rec-feas:cost}\\
        \textrm{s.t. } \, & \forall\, i=0, \mydots, N\!-\!1\!: \nonumber \\
        & z_0 = x(k), \label{SLTMPC-rec-feas:init}\\
        & z_{i+1} = Az_i + Bv_i + p_i, \label{SLTMPC-rec-feas:nom-dyn}\\
        & \begin{bmatrix} \mathbb{I}_{Nn} \!-\! \mathbf{ZA}_N & \ \scalebox{0.85}[1.0]{$-$}\mathbf{ZB}_N \end{bmatrix} \!\begin{bmatrix} \bm{\Phi}^\mathbf{e} \\ \bm{\Phi}^{\bm{\nu}} \end{bmatrix} = \bm{\Sigma}, \label{SLTMPC-rec-feas:error-dyn}\\
        & \Phi^{e}_{N,j-1} = A\Phi^{e}_{N,j} + B\Phi^{\nu}_{N,j} + \Xi_j, \label{SLTMPC-rec-feas:Toeplizt-constraint}\\[0.2cm]
        & z_i \in \mathcal{X} \ominus \mathcal{F}_i\left( \bm{\Phi}^\mathbf{e} \right), \label{SLTMPC-rec-feas:state-constraint}\\
        & v_i \in \mathcal{U} \ominus \mathcal{F}_i\left( \bm{\Phi}^{\bm{\nu}} \right), \label{SLTMPC-rec-feas:input-constraint}\\[0.2cm]
        & z_N \in \alpha\mathcal{Z}_f \subseteq \mathcal{X} \ominus \mathcal{F}_N\left( \bm{\Phi}^\mathbf{e} \right), \label{SLTMPC-rec-feas:terminal-constraint}\\
        & \alpha K_f\mathcal{Z}_f \subseteq \mathcal{U} \ominus \mathcal{F}_N\left( \bm{\Phi}^{\bm{\nu}} \right), \label{SLTMPC-rec-feas:terminal-input-constraint} \\
        & \alpha A_{K_f}\mathcal{Z}_f \subseteq \alpha \mathcal{Z}_f \ominus \Gamma \bar{\mathcal{W}}, \label{SLTMPC-rec-feas:terminal-decrease}\\[0.2cm]
        & \forall\:\! d=1, \mydots, n_D: \nonumber \\
        & \{ \psi^d_i \} \oplus \bigoplus_{j=0}^{i-1} \Psi^d_{i,j} \bar{\mathcal{W}} \oplus \mathcal{W} \subseteq \sigma_{i+1}\bar{\mathcal{W}}, \label{SLTMPC-rec-feas:dist-incl}\\
        & \alpha \Delta^d_{K_f} \mathcal{Z}_f \!\oplus\! \bigoplus_{j=0}^{N-1} \!\Psi^d_{N,j} \bar{\mathcal{W}} \!\oplus\! \mathcal{W} \subseteq \sigma_{N}\bar{\mathcal{W}}, \label{SLTMPC-rec-feas:terminal-dist-incl}
    \end{align}
\end{subequations}
where $l(\cdot, \cdot)$ and $l_f(\cdot)$ are stage and terminal costs according to Assumption~\ref{assump:cost}, $\psi^d_i$ and $\Psi^d_{i,j}$ are defined as in~\eqref{eq:q_Pi}, $\Psi^d_{N,j} = \Delta_A^d \Phi^{e}_{N,j} + \Delta_B^d \Phi^{\nu}_{N,j} - \Xi_{j}$, $\mathcal{Z}_f$ is an RPI set according to Definition~\ref{def:RPI}, and the resulting MPC control law is given by
\begin{equation}\label{eq:SLTMPC-rec-feas-law}
        \kappa_\textrm{MPC}(x(k)) = v_0^*,
\end{equation}
where $v_0^*$ is the first element of the optimal solution $\mathbf{v}^*$ of~\eqref{SLTMPC:rec-feas}. Details on how to reformulate \eqref{SLTMPC-rec-feas:state-constraint} - \eqref{SLTMPC-rec-feas:terminal-dist-incl} as linear constraints are provided in Appendix~\ref{app:implementation}.

\subsection{Theoretical Guarantees}
In the following, we prove recursive feasibility and input-to-state stability (ISS) of the proposed filter-based SLTMPC scheme~\eqref{SLTMPC:rec-feas}. To do this, we first discuss the equivalence of~\eqref{eq:dynamics} and~\eqref{eq:auxiliary-dynamics} by considering dynamics~\eqref{eq:dynamics} with the filter-based SLTMPC controller~\eqref{eq:SLTMPC-rec-feas-law}. The auxiliary disturbance~$\bar{w}(k)$ can be computed from the states measured at timesteps~$k, \, k\!+\!1$, the applied control input, and the disturbance filter variables~$p_0^*, \, \sigma_1^*$ as
\begin{equation}\label{eq:equivalent-disturbance}
        \bar{w}(k) = \frac{1}{\sigma_1^*}(x(k+1) - Ax(k) - B\kappa_\textrm{MPC}(x(k)) - p_0^*).
\end{equation}
Therefore, dynamics~\eqref{eq:dynamics} can be equivalently rewritten as
\begin{equation}\label{eq:equivalent-dynamics}
        x(k+1) = Ax(k) + B\kappa_\textrm{MPC}(x(k))  + p_0^* + \sigma_1^* \bar{w}(k).
\end{equation}
Using this equivalence, we show recursive feasibility of~\eqref{SLTMPC:rec-feas} in the following theorem.
\begin{thm}\label{prop:rec-feas}
Optimization problem~\eqref{SLTMPC:rec-feas} is recursively feasible for system~\eqref{eq:dynamics} under controller~\eqref{eq:SLTMPC-rec-feas-law}, i.e., the feasible set of~\eqref{SLTMPC:rec-feas} is RPI.
\end{thm}
\begin{pf}
We prove the theorem using a standard argument~\cite{Rawlings2009}, constructing a feasible candidate sequence based on the shifted previous solution, which is possible contrary to~\cite{Chen2023} because of the novel terminal constraints. We start the proof by stating constraint~\eqref{SLTMPC-rec-feas:error-dyn} in elementwise fashion, i.e.,
\begin{subequations}
\begin{alignat}{2}
\Phi^{e}_{i,i-1} &= \Sigma_{i,i-1} = \sigma_i \cdot \mathbb{I}_{n}, \quad&& i=1, \mydots, N, \label{proof:SLP-dyn-init} \\
\Phi^{e}_{i+1,j} &= A\Phi^{e}_{i,j} \!+\! B\Phi^{\nu}_{i,j} \!+\! \Sigma_{i+1,j},\ \: && i=1, \mydots, N\!-\!1, \nonumber \\
& && \!j=0, \mydots, i-1. \label{proof:SLP-dyn}
\end{alignat}
\end{subequations}
With these and the computed auxiliary disturbance~$\bar{w}(k)$ given by~\eqref{eq:equivalent-disturbance}, we define the scalar-valued candidate as $\hat{\alpha} = \alpha^*$, the vector-valued candidates as
\begin{subequations}\label{proof:vec-cand}
\begin{alignat}{2}
\hat{z}_i &= z_{i+1}^* + \Phi^{e\:\!*}_{i+1,0} \bar{w}(k), &&\quad i=0,\mydots,N\!-\!1, \label{proof:state-cand}\\
\hat{z}_N &= A_{K_f}z_N^* + \Gamma^* \bar{w}(k),\\
\hat{v}_i &= v_{i+1}^* + \Phi^{\nu\:\!*}_{i+1,0} \bar{w}(k), &&\quad i=0,\mydots,N\!-\!2,\\
\hat{v}_{N-1} &= K_f z_N^* + \Phi^{\nu\:\!*}_{N,0} \bar{w}(k), \\
\hat{p}_i &= p_{i+1}^* + \Sigma^*_{i+2,0} \bar{w}(k), &&\quad i=0,\mydots,N\!-\!2,\\
\hat{p}_{N-1} &= \Xi_{0}^* \bar{w}(k),
\end{alignat}
\end{subequations}
and the matrix-valued candidates as
\begin{subequations}\label{proof:mat-cand}
\begin{align}
\hat{\bm{\Phi}}^\mathbf{e} &= \mathrm{shift}(\bm{\Phi}^{\mathbf{e}\:\!*}, \bm{\Phi}^{\mathbf{e}\:\!*}_{N,:}),\\
\hat{\bm{\Phi}}^{\bm{\nu}} &= \mathrm{shift}(\bm{\Phi}^{\bm{\nu}\:\!*}, \bm{\Phi}^{\bm{\nu}\:\!*}_{N,:}),\\
\hat{\bm{\Sigma}} &= \mathrm{shift}(\bm{\Sigma}^{*}, [ \bm{\Xi}^{*}_{1:} \ \sigma_N^* \cdot \mathbb{I}_n ]),\\
\hat{\bm{\Xi}} &= \bm{\Xi}^{*},
\end{align}
\end{subequations}
where the $\mathrm{shift}(\cdot,\cdot)$ operator is defined in Section~\ref{sec:notation}, $\bm{\Phi}_{N,:}$ denotes the last block row of $\bm{\Phi}^\mathbf{e}$ and $\bm{\Phi}^{\bm{\nu}}$, and $\bm{\Xi}^{*}_{1:} = [ \Xi_{1}^* \ \mydots \ \Xi_{N-1}^* ]$. The matrix-valued candidates are thus constructed by shifting the optimal solutions up and left by one block row and block column, respectively, and by appending the last block row of $\bm{\Phi}^\mathbf{e}, \, \bm{\Phi}^{\bm{\nu}}$, and $[ \Xi_{1}^* \ \mydots \ \Xi_{N-1}^* \ \sigma_N^* \cdot \mathbb{I}_n ]$, respectively. Note that due to the choice of these candidates, we get $\hat{\sigma}_i = \sigma_{i+1}^*, \, i=0, \mydots, N\!-\!1, \, \hat{\sigma}_N = \sigma_N^*$, and $\hat{\Gamma} \!=\! \Gamma^* \!=\! A\Phi^{e\:\!*}_{N,0} + B\Phi^{\nu\:\!*}_{N,0} + \Xi_{0}^*$ by definition. Next, we show that candidates~\eqref{proof:vec-cand} fulfill nominal dynamics~\eqref{SLTMPC-rec-feas:nom-dyn}, error dynamics~\eqref{SLTMPC-rec-feas:error-dyn}, and the system response constraint~\eqref{SLTMPC-rec-feas:Toeplizt-constraint}. Using~\eqref{proof:state-cand} with $i=0$, the state candidate sequence is initialized with
\begin{align*}
\hat{z}_0 &= z_1^* + \Phi^{e\:\!*}_{1,0} \bar{w}(k) = Az_0^* + Bv_0^* + p_0^* + \sigma_1^* \bar{w}(k) \\
&= Ax(k) + B\kappa_\textrm{MPC}(x(k)) + \eta(k) = x(k+1),
\end{align*}
where we used~$\Phi^{e\:\!*}_{1,0} = \sigma_1^* \cdot \mathbb{I}_{n}$ by~\eqref{proof:SLP-dyn-init}, nominal dynamics~\eqref{SLTMPC-rec-feas:nom-dyn}, and~\eqref{eq:equivalent-dynamics}. For the remainder of the state candidate sequence, i.e., $i=1, \mydots, N-1$, it holds that
\begin{align*}
\hat{z}_{i} &= z_{i+1}^* + \Phi^{e\:\!*}_{i+1,0} \bar{w}(k) \\
&\overset{\mathclap{\strut\text{$\substack{\eqref{SLTMPC-rec-feas:nom-dyn}\\ \eqref{proof:SLP-dyn}}$}}}= Az_{i}^* + Bv_{i}^* + p_{i}^* + \left(A\Phi^{e\:\!*}_{i,0} + B\Phi^{\nu\:\!*}_{i,0} + \Sigma_{i+1,0}^*\right)\bar{w}(k) \\
&= A\hat{z}_{i-1} + B\hat{v}_{i-1} + \hat{p}_{i-1},
\end{align*}
and for the last state in the candidate sequence, i.e. $i=N$, we get
\begin{align*}
\hat{z}_{N}\! &= \!A_{K_f}z_N^* + \Gamma^*\bar{w}(k) \\
&= \!A_{K_f}z_{N}^* + \left(A\Phi^{e\:\!*}_{N,0} + B\Phi^{\nu\:\!*}_{N,0} + \Xi_{0}^*\right)\bar{w}(k) \\
&= \!A(z_N^* \!+\! \Phi^{e\:\!*}_{N,0}\bar{w}(k)) \!+\! B(K_fz_N^* \!+\! \Phi^{\nu\:\!*}_{N,0}\bar{w}(k)) \!+\! \Xi_{0}^*\bar{w}(k) \\
&= \!A\hat{z}_{N-1} + B\hat{v}_{N-1} + \hat{p}_{N-1},
\end{align*}
by definition of $\Gamma^*$ and candidate sequences~\eqref{proof:vec-cand}, which shows that the candidate sequences fulfill the nominal dynamics. For the error dynamics, the block diagonal of~$\hat{\bm{\Phi}}^\mathbf{e}$ is initialized for $i=1, \mydots, N\!-\!1$ as
\begin{align*}
\hat{\Phi}^{e}_{i,i-1} &= \Phi^{e\:\!*}_{i+1,i} = \sigma_{i+1}^* \cdot \mathbb{I}_{n} = \hat{\sigma}_i \cdot \mathbb{I}_{n},
\end{align*}
and for $i=N$ as
\begin{equation*}
\hat{\Phi}^{e}_{N,N-1} = \Phi^{e\:\!*}_{N,N-1} = \sigma_{N}^* \cdot \mathbb{I}_{n} = \hat{\sigma}_N \cdot \mathbb{I}_{n}.
\end{equation*}
Then, for $i=2, \mydots, N\!-\!1$ and $j=0, \mydots, i\!-\!2$ we get
\begin{align*}
\hat{\Phi}^{e}_{i,j} &= \Phi^{e\:\!*}_{i+1,j+1} \, \overset{\mathclap{\strut\text{\eqref{proof:SLP-dyn}}}}= \, A\Phi^{e\:\!*}_{i,j+1} + B\Phi^{\nu\:\!*}_{i,j+1} + \Sigma^*_{i+1,j+1} \\
&= A\hat{\Phi}^{e}_{i-1,j} + B\hat{\Phi}^{\nu}_{i-1,j} + \hat{\Sigma}_{i,j},
\end{align*}
and for $i=N$ and $j=0, \mydots, N-2$, we get
\begin{align*}
\hat{\Phi}^{e}_{N,j} &= \Phi^{e\:\!*}_{N,j} \, \overset{\mathclap{\strut\text{\eqref{SLTMPC-rec-feas:Toeplizt-constraint}}}}= \, A\Phi^{e\:\!*}_{N,j+1} + B\Phi^{\nu\:\!*}_{N,j+1} + \Xi_{j+1}^* \\
&= A\hat{\Phi}^{e}_{N-1,j} + B\hat{\Phi}^{\nu}_{N-1,j} + \hat{\Sigma}_{N,j}.
\end{align*}
Constraint~\eqref{SLTMPC-rec-feas:Toeplizt-constraint} is trivially fulfilled, i.e.,
\begin{align*}
\hat{\Phi}^{e}_{N,j-1} &= \Phi^{e\:\!*}_{N,j-1} = A\Phi^{e\:\!*}_{N,j} + B\Phi^{\nu\:\!*}_{N,j} + \Xi_{j}^* \\
&= A\hat{\Phi}^{e}_{N,j} + B\hat{\Phi}^{\nu}_{N,j} + \hat{\Xi}_j.
\end{align*}
Next, we need to show that candidate sequences~\eqref{proof:vec-cand} fulfill the inclusions~\eqref{SLTMPC-rec-feas:state-constraint} and~\eqref{SLTMPC-rec-feas:input-constraint}. For this, we first define the tube sequences via the candidate matrices, i.e., for $i=0, \mydots, N\!-\!1$:
\begin{align*}
\mathcal{F}_i(\hat{\bm{\Phi}}^\mathbf{e}) &= \bigoplus_{j=0}^{i-1} \hat{\Phi}^{e}_{i,j} \bar{\mathcal{W}} = \bigoplus_{j=1}^{i} \Phi^{e\:\!*}_{i+1,j} \bar{\mathcal{W}}, \\
\mathcal{F}_i(\hat{\bm{\Phi}}^{\bm{\nu}}) &= \bigoplus_{j=0}^{i-1} \hat{\Phi}^{\nu}_{i,j} \bar{\mathcal{W}} = \bigoplus_{j=1}^{i} \Phi^{\nu\:\!*}_{i+1,j} \bar{\mathcal{W}}.
\end{align*}
Note that above set definitions together with~\eqref{eq:tubes} yield the following relations
\begin{align*}
\mathcal{F}_{i+1}(\bm{\Phi}^{\mathbf{e}\:\!*}) &= \mathcal{F}_{i}(\hat{\bm{\Phi}}^\mathbf{e}) \oplus \Phi^{e\:\!*}_{i+1,0}\bar{\mathcal{W}}, \\
\mathcal{F}_{i+1}(\bm{\Phi}^{\bm{\nu}\:\!*}) &= \mathcal{F}_{i}(\hat{\bm{\Phi}}^{\bm{\nu}}) \oplus \Phi^{\nu\:\!*}_{i+1,0}\bar{\mathcal{W}}.
\end{align*}
To prove that the inclusion constraints are fulfilled, we need to show that for $i=0, \mydots, N\!-\!1$
\begin{subequations}\label{proof:constraint-satisfaction}
\begin{align}
\hat{z}_i = z_{i+1}^* \!+  \Phi^{e\:\!*}_{i+1,0} \bar{w}(k) &\in \mathcal{X} \ominus \mathcal{F}_i(\hat{\bm{\Phi}}^\mathbf{e}), \label{proof:state-cont}\\
\hat{v}_i = v_{i+1}^* \!+ \Phi^{\nu\:\!*}_{i+1,0} \bar{w}(k) &\in \mathcal{U} \ominus \mathcal{F}_i(\hat{\bm{\Phi}}^{\bm{\nu}}). \label{proof:input-cont}
\end{align}
\end{subequations}
Given that $z_{i+1}^* \in \mathcal{X} \ominus \mathcal{F}_{i+1}(\bm{\Phi}^{\mathbf{e}\:\!*})$ and $\Phi^{e\:\!*}_{i+1,0} \bar{w}(k) \in \Phi^{e\:\!*}_{i+1,0}\bar{\mathcal{W}}$, inclusion~\eqref{proof:state-cont} holds for $i=0,\mydots,N\!-\!2$ with
\begin{align*}
&\left(\mathcal{X} \ominus \mathcal{F}_{i+1}(\bm{\Phi}^{\mathbf{e}\:\!*}) \right) \oplus \Phi^{e\:\!*}_{i+1,0}\bar{\mathcal{W}} \\
&= \left(\mathcal{X} \!\ominus\! \mathcal{F}_{i}(\hat{\bm{\Phi}}^\mathbf{e}) \!\ominus \Phi^{e\:\!*}_{i+1,0}\bar{\mathcal{W}} \right) \!\oplus \Phi^{e\:\!*}_{i+1,0}\bar{\mathcal{W}} \subseteq \mathcal{X} \!\ominus\! \mathcal{F}_i(\hat{\bm{\Phi}}^\mathbf{e}),
\end{align*}
where we used~\cite[Theorem~2.1~(ii), (v)]{Kolmanovsky1998} and similarly inclusion~\eqref{proof:input-cont} holds for $i=0,\mydots,N\!-\!2$ with
\begin{align*}
&\left(\mathcal{U} \ominus \mathcal{F}_{i+1}(\bm{\Phi}^{\bm{\nu}\:\!*}) \right) \oplus \Phi^{\nu\:\!*}_{i+1,0}\bar{\mathcal{W}} \\
&= \left(\mathcal{U} \!\ominus\! \mathcal{F}_{i}(\hat{\bm{\Phi}}^{\bm{\nu}}) \!\ominus \Phi^{\nu\:\!*}_{i+1,0}\bar{\mathcal{W}} \right) \!\oplus \Phi^{\nu\:\!*}_{i+1,0}\bar{\mathcal{W}} \subseteq \mathcal{U} \!\ominus\! \mathcal{F}_i(\hat{\bm{\Phi}}^{\bm{\nu}}).
\end{align*}
For the last candidate state and input, we need to show
\begin{align*}
\hat{z}_{N-1} &= z_{N}^* +  \Phi^{e\:\!*}_{N,0} \bar{w}(k) \in \mathcal{X} \ominus \mathcal{F}_{N-1}(\hat{\bm{\Phi}}^\mathbf{e}),\\
\hat{v}_{N-1} &= K_f z_{N}^* + \Phi^{\nu\:\!*}_{N,0} \bar{w}(k) \in \mathcal{U} \ominus \mathcal{F}_{N-1}(\hat{\bm{\Phi}}^{\bm{\nu}}).
\end{align*}
Given that $z_{N}^* \in \alpha^*\mathcal{Z}_f \subseteq \mathcal{X} \ominus \mathcal{F}_{N}(\bm{\Phi}^{\mathbf{e}\:\!*})$, $\Phi^{e\:\!*}_{N,0} \bar{w}(k) \in \Phi^{e\:\!*}_{N,0}\bar{\mathcal{W}}$, $K_f z_{N}^* \in \alpha^* K_f\mathcal{Z}_f \subseteq \mathcal{U} \ominus \mathcal{F}_{N}(\bm{\Phi}^{\bm{\nu}\:\!*})$, and $\Phi^{\nu\:\!*}_{N,0} \bar{w}(k) \in \Phi^{\nu\:\!*}_{N,0}\bar{\mathcal{W}}$, we can confirm that
\begin{align*}
\left(\mathcal{X} \ominus \mathcal{F}_{N}(\bm{\Phi}^{\mathbf{e}\:\!*}) \right) \oplus \Phi^{e\:\!*}_{N,0}\bar{\mathcal{W}} &\subseteq \mathcal{X} \ominus \mathcal{F}_{N-1}(\hat{\bm{\Phi}}^\mathbf{e}), \\
\left(\mathcal{U} \ominus \mathcal{F}_{N}(\bm{\Phi}^{\bm{\nu}\:\!*}) \right) \oplus \Phi^{\nu\:\!*}_{N,0}\bar{\mathcal{W}} &\subseteq \mathcal{U} \ominus \mathcal{F}_{N-1}(\hat{\bm{\Phi}}^{\bm{\nu}}).
\end{align*}
The terminal set constraints~\eqref{SLTMPC-rec-feas:terminal-constraint}-\eqref{SLTMPC-rec-feas:terminal-decrease} are trivially fulfilled, since the last block row of all candidate matrices is equal to the last block row of the previous optimal solution (see~\eqref{proof:mat-cand}) and $\hat{\alpha} = \alpha^*$, therefore we get
\begin{align*}
\hat{\alpha} A_{K_f}\mathcal{Z}_f \!\subseteq\! \hat{\alpha} \mathcal{Z}_f \!\ominus\! \hat{\Gamma} \bar{\mathcal{W}} &\Leftrightarrow \alpha^* \!A_{K_f}\mathcal{Z}_f \!\subseteq\! \alpha^* \mathcal{Z}_f \!\ominus\! \Gamma^* \bar{\mathcal{W}}, \\
\hat{\alpha} \mathcal{Z}_f \!\subseteq\! \mathcal{X} \!\ominus\! \mathcal{F}_N( \hat{\bm{\Phi}}^\mathbf{e} ) &\Leftrightarrow \alpha^* \!\mathcal{Z}_f \!\subseteq\! \mathcal{X} \!\ominus\! \mathcal{F}_N( \bm{\Phi}^{\mathbf{e}\:\!*} ), \\
\hat{\alpha} K_f\mathcal{Z}_f \!\subseteq \mathcal{U} \!\ominus\! \mathcal{F}_N( \hat{\bm{\Phi}}^{\bm{\nu}} ) &\Leftrightarrow \alpha^* \!K_f\mathcal{Z}_f \!\subseteq \mathcal{U} \!\ominus\! \mathcal{F}_N( \bm{\Phi}^{\bm{\nu}\:\!*} ).
\end{align*}
The terminal constraint $\hat{z}_N \in \hat{\alpha}\mathcal{Z}_f$ in~\eqref{SLTMPC-rec-feas:terminal-constraint} is also trivially fulfilled since
\begin{equation*}
\hat{z}_N \!=\! A_{K_f}z_N^* + \Gamma^* \bar{w}(k) \!\in\! \alpha^*\!A_{K_f}\mathcal{Z}_f \oplus \Gamma^* \bar{\mathcal{W}} \overset{\mathclap{\strut\text{\,\eqref{SLTMPC-rec-feas:terminal-decrease}}}}\subseteq \alpha^*\!\mathcal{Z}_f \!=\! \hat{\alpha}\mathcal{Z}_f.
\end{equation*}
To show that~\eqref{SLTMPC-rec-feas:dist-incl} and~\eqref{SLTMPC-rec-feas:terminal-dist-incl} hold, we first define the candidates of auxiliary variables $\psi^d_i,\, \Psi^d_{i,j}$ for a single vertex~$\mathcal{D}^d$, i.e., for $i=0, \mydots, N\!-\!2$ we have
\begin{align*}
\hat{\psi}^d_i &= \Delta_A^d \hat{z}_i + \Delta_B^d \hat{v}_i - \hat{p}_i \\
&= \Delta_A^d z_{i+1}^* + \Delta_B^d v_{i+1}^* - p_{i+1}^* \\
&\quad+ \left(\Delta_A^d \Phi^{e\:\!*}_{i+1,0} + \Delta_B^d \Phi^{\nu\:\!*}_{i+1,0} - \Sigma_{i+2,0}^*\right) \bar{w}(k) \\
&= \psi_{i+1}^{d\:\!*} + \Psi^{d\:\!*}_{i+1,0}\bar{w}(k), \\
\hat{\Psi}^{d}_{i,j} &= \Delta_A^d \hat{\Phi}^{e}_{i,j} + \Delta_B^d \hat{\Phi}^{\nu}_{i,j} - \hat{\Sigma}_{i+1,j} \\
&= \Delta_A^d \Phi^{e\:\!*}_{i+1,j+1} + \Delta_B^d \Phi^{\nu\:\!*}_{i+1,j+1} - \Sigma_{i+2,j+1}^* \\
&= \Psi^{d\:\!*}_{i+1, j+1},
\end{align*}
and for the last timesteps, we get
\begin{align*}
\hat{\psi}^d_{N-1} &= \Delta_A^d \hat{z}_{N-1} + \Delta_B^d \hat{v}_{N-1} - \hat{p}_{N-1} \\
&= \Delta_A^d z_{N}^* + \Delta_B^d K_f z_{N}^* \\
&\quad+ \left(\Delta_A^d \Phi^{e\:\!*}_{N,0} + \Delta_B^d \Phi^{\nu\:\!*}_{N,0} - \Xi_{0}^*\right) \bar{w}(k) \\
&= \Delta^d_{K_f} z_N^* + \Psi^{d\:\!*}_{N,0}\bar{w}(k), \\
\hat{\Psi}^{d}_{N-1,j} &= \Delta_A^d \hat{\Phi}^{e}_{N-1,j} + \Delta_B^d \hat{\Phi}^{\nu}_{N-1,j} - \hat{\Sigma}_{N,j} \\
&= \Delta_A^d \Phi^{e\:\!*}_{N,j+1} + \Delta_B^d \Phi^{\nu\:\!*}_{N,j+1} - \Xi_{j+1}^* = \Psi^{d\:\!*}_{N, j+1}, \\
\hat{\Psi}^d_{N,j} &= \Delta_A^d \hat{\Phi}^{e}_{N,j} + \Delta_B^d \hat{\Phi}^{\nu}_{N,j} - \hat{\Xi}_{j} \\
&= \Delta_A^d \Phi^{e\:\!*}_{N,j} + \Delta_B^d \Phi^{\nu\:\!*}_{N,j} - \Xi_{j}^* = \Psi^{d\:\!*}_{N,j}.
\end{align*}
Using these candidates, we then check the inclusions~\eqref{SLTMPC-rec-feas:dist-incl} and~\eqref{SLTMPC-rec-feas:terminal-dist-incl}, i.e., for $i=0, \mydots, N\!-\!2$ we get
\begin{align*}
& \{ \hat{\psi}^d_i \} \oplus \bigoplus_{j=0}^{i-1} \hat{\Psi}^d_{i,j} \bar{\mathcal{W}} \oplus \mathcal{W} \\
&= \{ \psi_{i+1}^{d\:\!*} + \Psi^{d\:\!*}_{i+1,0} \bar{w}(k) \} \oplus \bigoplus_{j=1}^{i} \Psi^{d\:\!*}_{i+1,j} \bar{\mathcal{W}} \oplus \mathcal{W} \\
&\subseteq \{ \psi_{i+1}^{d\:\!*} \} \oplus \bigoplus_{j=0}^{i} \Psi^{d\:\!*}_{i+1,j} \bar{\mathcal{W}} \oplus \mathcal{W} \\
&\overset{\mathclap{\strut\text{\,\eqref{SLTMPC-rec-feas:dist-incl}}}}\subseteq \sigma_{i+2}^* \bar{\mathcal{W}} = \hat{\sigma}_{i+1} \bar{\mathcal{W}},
\end{align*}
and for $i=N\!-\!1$ we get
\begin{align*}
& \{ \hat{\psi}^d_{N-1} \} \oplus \bigoplus_{j=0}^{N-2} \hat{\Psi}^d_{N-1,j} \bar{\mathcal{W}} \oplus \mathcal{W} \\
&= \{ \Delta^d_{K_f} z_{N}^* + \Psi^{d\:\!*}_{N,0} \bar{w}(k) \} \oplus \bigoplus_{j=1}^{N-1} \Psi^{d\:\!*}_{N,j} \bar{\mathcal{W}} \oplus \mathcal{W} \\
&\subseteq \alpha^* \Delta^d_{K_f} \mathcal{Z}_f \oplus \bigoplus_{j=0}^{N-1} \Psi^{d\:\!*}_{N,j} \bar{\mathcal{W}} \oplus \mathcal{W} \\
&\overset{\mathclap{\strut\text{\,\eqref{SLTMPC-rec-feas:terminal-dist-incl}}}}\subseteq \sigma_{N}^* \bar{\mathcal{W}} = \hat{\sigma}_{N} \bar{\mathcal{W}},
\end{align*}
both of which satisfy~\eqref{SLTMPC-rec-feas:dist-incl}. Finally,~\eqref{SLTMPC-rec-feas:terminal-dist-incl} is trivially fulfilled by the candidate solution, i.e.,
\begin{align*}
&\hat{\alpha} \Delta^d_{K_f} \mathcal{Z}_f \oplus \bigoplus_{j=0}^{N-1} \hat{\Psi}^d_{N,j} \bar{\mathcal{W}} \oplus \mathcal{W} \\
&= \alpha^* \Delta^d_{K_f} \mathcal{Z}_f \oplus \bigoplus_{j=0}^{N-1} \Psi^{d\:\!*}_{N,j} \bar{\mathcal{W}} \oplus \mathcal{W} \\
&\overset{\mathclap{\strut\text{\,\eqref{SLTMPC-rec-feas:terminal-dist-incl}}}}\subseteq \sigma_{N}^*\bar{\mathcal{W}} = \hat{\sigma}_{N}\bar{\mathcal{W}}.
\end{align*}
Since we impose~\eqref{SLTMPC-rec-feas:dist-incl} and~\eqref{SLTMPC-rec-feas:terminal-dist-incl} for each vertex $\mathcal{D}^d$ individually, the argument holds for all vertices of~$\mathcal{D}$. Therefore, we have shown that the candidate solutions~\eqref{proof:vec-cand},~\eqref{proof:mat-cand} satisfy all constraints in~\eqref{SLTMPC:rec-feas}, which proves recursive feasibility. \qed
\end{pf}
Next, we show that the closed-loop system is input-to-state stable~(ISS), where ISS is formally defined in~\cite[Definition~B.42]{Rawlings2009}. For the ISS proof, we rely on an ISS-Lyapunov function~\cite[Appendix~B.6]{Rawlings2009}, which we define in the following definition. Note that due to the equivalence of~\eqref{eq:dynamics} and~\eqref{eq:equivalent-dynamics}, we only need to check the stability of~\eqref{eq:equivalent-dynamics} to conclude ISS of the original system.
\begin{definition}[ISS-Lyapunov function]\label{def:ISS-Lyap}
Consider an RPI set~$\mathcal{S} \subset \mathbb{R}^n$ with the origin in its interior. A function $V: \mathcal{S} \to \mathbb{R}_{\geq 0}$ is called an ISS-Lyapunov function in~$\mathcal{S}$ for system~\eqref{eq:equivalent-dynamics} with $u = \kappa_\textrm{MPC}(x)$ and~$\bar{w} \in \bar{\mathcal{W}}$, if there exist $\mathcal{K}$-functions~$\gamma_1, \gamma_2$, and~$\gamma_3$, such that for all $x \in \mathcal{S}$ and $\bar{w} \in \bar{\mathcal{W}}$:
\begin{subequations}
\begin{align}
&\gamma_1(\|x\|) \leq V(x) \leq \gamma_2(\|x\|), \label{ISS-L:bounds}\\
&V(x^+) \!-\! V(x) \leq \!-\gamma_1(\|x\|) \!+\! \gamma_3(\|\bar{w}\|), \label{ISS-L:decrease}
\end{align}
\end{subequations}
where~$\| \cdot \|$ is any vector norm on~$\mathbb{R}^n$.
\end{definition}
\begin{thm}\label{prop:ISS}
        Given that Assumption~\ref{assump:cost} holds, system~\eqref{eq:dynamics} subject to constraints~\eqref{eq:constraints} and in closed-loop with~\eqref{eq:SLTMPC-rec-feas-law}, is input-to-state stable (ISS) in $\mathcal{X}_\textrm{feas}$ for any admissible sequence of combined disturbances $\bm{\eta}$, where $\mathcal{X}_\textrm{feas}$ is the set of all initial states $x(k)$ for which~\eqref{SLTMPC:rec-feas} is feasible.
\end{thm}
\begin{pf}
        Due to Theorem~\ref{prop:rec-feas}, we know that~\eqref{SLTMPC:rec-feas} is recursively feasible for all $x(k) \in \mathcal{X}_\textrm{feas}$. Therefore, we can prove ISS by showing that the optimal value function of~\eqref{SLTMPC:rec-feas} is an ISS-Lyapunov function in $\mathcal{X}_\textrm{feas}$ according to Definition~\ref{def:ISS-Lyap}~\cite[Lemma~B.44]{Rawlings2009}. Denote~\eqref{SLTMPC-rec-feas:cost} as $V_N(x(k);\mathbf{z},\mathbf{v})$ and the optimal value function of~\eqref{SLTMPC:rec-feas} as $V_N^*(x(k))$, i.e., $V_N(\cdot)$ evaluated at the optimal solution $\mathbf{z}^*, \mathbf{v}^*$. Using the candidate sequences $\hat{\mathbf{z}}, \hat{\mathbf{v}}$ defined in~\eqref{proof:vec-cand}, it follows that
        \begin{align*}
                V_N^*(x(k\!+\!1)) \!-\! V_N^*(x(k)) &\leq V_N(x(k\!+\!1);\hat{\mathbf{z}}, \hat{\mathbf{v}}) \!-\! V_N^*(x(k)) \\
                &= l_f(\hat{z}_N) \!-\! l_f(z_N^*) \\
                &\quad+ \sum_{i=0}^{N-1} l(\hat{z}_i, \hat{v}_i) \!-\! l(z_i^*, v_i^*).
        \end{align*}
        We then use the candidate sequences~\eqref{proof:vec-cand} and uniform continuity of $l_f(\cdot)$ with corresponding $\mathcal{K}$-function $\beta_f(\cdot)$ to obtain the upper bound
        \begin{align*}
                l_f(\hat{z}_N) \!-\! l_f(z_N^*) &= l_f(A_{K_f}z_{N}^* + \Gamma^*\bar{w}(k)) - l_f(z_N^*) \\
                &\quad +l_f(A_{K_f}z_{N}^*) - l_f(A_{K_f}z_{N}^*) \\
                &\leq l_f(A_{K_f}z_{N}^*) - l_f(z_N^*) \\
                &\quad\!+\! \| l_f(A_{K_f}z_{N}^* \!+\! \Gamma^*\bar{w}(k)) \!-\! l_f(A_{K_f}z_{N}^*)\| \\
                &\leq l_f(A_{K_f}z_{N}^*) - l_f(z_N^*) \!+\! \beta_f(\|\Gamma^*\bar{w}(k)\|).
        \end{align*}
        We rewrite the sum of the cost differences as
        \begin{align*}
                \sum_{i=0}^{N-1} l(\hat{z}_i, \hat{v}_i) - l(z_i^*, v_i^*) &= \scalebox{0.75}[1.0]{\( - \)} l(z_0^*, v_0^*) + l(\hat{z}_{N-1}, \hat{v}_{N-1}) \\
                &\quad +\!\!\sum_{i=1}^{N-1} l(\hat{z}_{i-1}, \hat{v}_{i-1}) - l(z_i^*, v_i^*),
        \end{align*}
        and use the candidate sequences~\eqref{proof:vec-cand} and uniform continuity of $l(\cdot, \cdot)$ with $\beta_x(\cdot), \beta_u(\cdot)$ the respective $\mathcal{K}$-functions to obtain
        \begin{align*}
                l(\hat{z}_{i\,\textrm{-}1}, \!\hat{v}_{i\,\textrm{-}1}\;\!\!) \!-\! l(z_i^*\!, v_i^*\;\!\!) \!&= l(z_{i}^* \!+\! \Phi^{e\:\!*}_{i,0}\bar{w}(k), v_{i}^* \!+\! \Phi^{\nu\:\!*}_{i,0}\bar{w}(k)) \\
                &\quad- l(z_i^*, v_i^*) \\
                &= l(z_{i}^* \!+\! \Phi^{e\:\!*}_{i,0}\bar{w}(k), v_{i}^* \!+\! \Phi^{\nu\:\!*}_{i,0}\bar{w}(k)) \\
                &\quad- l(z_{i}^*, v_{i}^* \!+\! \Phi^{\nu\:\!*}_{i,0}\bar{w}(k)) \\
                &\quad+ l(z_{i}^*, v_{i}^* \!+\! \Phi^{\nu\:\!*}_{i,0}\bar{w}(k)) - l(z_i^*, v_i^*) \\
                &\!\leq \!\beta_x(\|\Phi^{e\:\!*}_{i,0}\bar{w}(k)\|) \!+\! \beta_u(\|\Phi^{\nu\:\!*}_{i,0}\bar{w}(k)\|).
        \end{align*}
        Additionally, we rewrite
        \begin{align*}
                l(\hat{z}_{N-1}, \hat{v}_{N-1}) &= l(z_{N}^* \!+\! \Phi^{e\:\!*}_{N,0}\bar{w}(k), K_fz_{N}^* \!+\! \Phi^{\nu\:\!*}_{N,0}\bar{w}(k)) \\
                &= l(z_{N}^* \!+\! \Phi^{e\:\!*}_{N,0}\bar{w}(k), K_fz_{N}^* \!+\! \Phi^{\nu\:\!*}_{N,0}\bar{w}(k)) \\
                &\quad- l(z_{N}^*, K_fz_{N}^* \!+\! \Phi^{\nu\:\!*}_{N,0}\bar{w}(k)) \\
                &\quad+ l(z_{N}^*, K_fz_{N}^* \!+\! \Phi^{\nu\:\!*}_{N,0}\bar{w}(k)) \\
                &\quad - l(z_{N}^*, K_fz_{N}^*) + l(z_{N}^*, K_fz_{N}^*) \\
                &\leq \beta_x(\|\Phi^{e\:\!*}_{N,0}\bar{w}(k)\|) + \beta_u(\|\Phi^{\nu\:\!*}_{N,0}\bar{w}(k)\|) \\
                &\quad + l(z_{N}^*, K_fz_{N}^*),
        \end{align*}
        and by combining the above, we get the value function decrease
        \begin{align*}
                V_N^*(x(k\!+\!1)) \!-\! V_N^*(x(k)) &\leq - l(z_0^*, v_0^*) + l(z_{N}^*, K_fz_{N}^*) \\
                &\quad + l_f(A_{K_f}z_{N}^*) - l_f(z_N^*) \\
                &\quad +\gamma_3(\|\bar{w}(k)\|),
        \end{align*}
        where we collected all $\mathcal{K}$-functions in
        \begin{align}
                \gamma_3(\|\bar{w}(k)\|) &= \beta_f(\|\Gamma^*\bar{w}(k)\|) \label{proof:beta-tot}\\
                &\quad+ \!\sum_{i=1}^{N} \beta_x(\|\Phi^{e\:\!*}_{i,0}\bar{w}(k)\|) \!+\! \beta_u(\|\Phi^{\nu\:\!*}_{i,0}\bar{w}(k)\|). \nonumber
        \end{align}
        Due to compactness of~$\mathcal{X}$ and~$\mathcal{U}$, the variables $\Phi^{e\:\!*}_{i,0}, \, \Phi^{\nu\:\!*}_{i,0}$, and $\Gamma^*$ are bounded. Therefore,~\eqref{proof:beta-tot} is a sum of compositions of $\mathcal{K}$-functions, e.g. $\beta_f(\|\Gamma^*\bar{w}(k)\|) \leq \beta_f(\|\Gamma^*\| \|\bar{w}(k)\|)$, which we can rewrite as a single $\mathcal{K}$-function $\gamma_3(\cdot)$. Finally, we notice that $-l(z_0^*, v_0^*) = -l(x(k), \kappa_\textrm{MPC}(x(k))) \leq -\gamma_1(\|x(k)\|)$ and $l_f(A_{K_f}z_N^* ) - l_f(z_N^*) \leq -l(z_{N}^*, K_fz_{N}^*)$ due to Assumption~\ref{assump:cost} and therefore obtain
        \begin{align*}
                V_N^*(x(k\!+\!1)) \!-\! V_N^*(x(k)) &\leq -\gamma_1(\|x(k)\|) \!+\! \gamma_3(\|\bar{w}(k)\|).
        \end{align*}
        Since $\gamma_3(\cdot)$ is a $\mathcal{K}$-function, $V_N^*(\cdot)$ fulfills decrease condition~\eqref{ISS-L:decrease} and because $V_N^*(x) \geq l(x, \kappa_\textrm{MPC}(x))\, \forall x \in \mathcal{X}_\textrm{feas}$, the optimal value function is lower bounded by the $\mathcal{K}$-function $\gamma_1(\cdot)$ by assumption, i.e., $V_N^*(x) \geq \gamma_1(\|x\|)\, \forall x \in \mathcal{X}_\textrm{feas}$. Using the facts that $l_f(\cdot)$ is upper bounded, the state and input constraints are compact, and monotonicity of $V_N^*(\cdot)$, we can conclude that $V_N^*(\cdot)$ is upper bounded by a $\mathcal{K}$-function~$\gamma_2(\cdot)$ in $\mathcal{X}_\mathrm{feas}$ due to~\cite[Propositions 2.15 - 2.16]{Rawlings2009}. Therefore, $V_N^*(\cdot)$ also fulfills~\eqref{ISS-L:bounds} and is thus an ISS-Lyapunov function in $\mathcal{X}_\mathrm{feas}$ (Definition~\ref{def:ISS-Lyap}). Hence, we have shown ISS of system~\eqref{eq:equivalent-dynamics} in closed-loop with~\eqref{eq:SLTMPC-rec-feas-law}, which implies ISS of the original system~\eqref{eq:dynamics} with SLTMPC controller~\eqref{eq:SLTMPC-rec-feas-law}. \qed
\end{pf}
\vspace{-0.3cm}
\begin{remark}
        Contrary to~\cite{Limon2009,Raimondo2009}, we do not treat the components of the combined uncertainty $\eta = d + w$ separately, where the state dependent model uncertainty~$d$ is assumed to have a stability margin that is incorporated in~$\scalebox{0.7}[1.0]{\( - \)}\,\gamma_1(\|x\|)$~\cite[Remark~3]{Raimondo2009}. Instead we use~\eqref{eq:equivalent-dynamics} with a single disturbance and the standard ISS proof technique~\cite[Appendix~B.6]{Rawlings2009}. Intuitively, we treat~\eqref{eq:equivalent-dynamics} as\vspace{-0.1cm}
        \begin{equation*}
                x^+ = Ax + \begin{bmatrix} B & I_n \end{bmatrix}\begin{bmatrix} v_0^* \\ p_0^* \end{bmatrix} + \sigma_1^*w(k),
                \vspace{-0.1cm}
        \end{equation*}
        in the proof, where~$[v_0^{* \top} \ p_0^{* \top}]^\top$ is an extended input and~$\sigma_1^*$ is an optimized but bounded disturbance gain. Also note that~$\Phi^{e\:\!*}_{i,0}, \, \Phi^{\nu\:\!*}_{i,0}$, and $\Gamma^*$ in~\eqref{proof:beta-tot} are optimization variables that vary each time~\eqref{SLTMPC:rec-feas} is solved. Therefore, they affect the size of the region around the origin where the closed-loop system converges to. Adding a regularizer on these variables thus allows tuning of this convergence region.
\end{remark}

\begin{table*}
        \centering
        \caption{Computational complexity of the two proposed SLTMPC methods~\eqref{SLTMPC:generic},~\eqref{SLTMPC:rec-feas} and the method proposed in~\cite{Chen2023} in terms of optimization variables and constraints. The dimensions used below, e.g. $n_x$, are defined in Section~\ref{sec:preliminaries}.}\label{table:comp-complexity}
        \vspace*{0.15cm}
        \begin{tabular}{@{}lll@{}}
        \toprule
        & Number of variables & Number of constraints \\ \midrule
        Chen et al.~\cite{Chen2023} & $\begin{aligned}\mathcal{O}(&N^2(n^2 + nm + n_xn_{\tilde{w}} + n_u n_{\tilde{w}})\\ &+ N(n_f n_{\tilde{w}}))\end{aligned}$ & $\begin{aligned}\mathcal{O}(&N^2n_{\tilde{w}}(n_x + n_u)\\ &+ N(n + n_f n_{\tilde{w}}))\end{aligned}$ \\[0.4cm]
        SLTMPC~\eqref{SLTMPC:generic} & $\begin{aligned}\mathcal{O}(&N^2(n^2 + nm + n_xn_{\tilde{w}} + n_u n_{\tilde{w}})\\ &+ N(n_f n_{\tilde{w}} + n_{\tilde{w}}^2n_D))\end{aligned}$ & $\begin{aligned}\mathcal{O}(&N^2n_{\tilde{w}}(n_x + n_u) \\ &+ N(n + n_f n_{\tilde{w}} + n_{\tilde{w}}^2 n_D))\end{aligned}$ \\[0.4cm]
        Rec. Feas. SLTMPC~\eqref{SLTMPC:rec-feas} & $\begin{aligned}\mathcal{O}(&N^2(n^2 + nm + n_xn_{\tilde{w}} + n_u n_{\tilde{w}})\\ &+ N(n_f n_{\tilde{w}} + n_{\tilde{w}}^2n_D + n^2) + n_f^2)\end{aligned}$ & $\begin{aligned}\mathcal{O}(&N^2n_{\tilde{w}}(n_x + n_u )\\ &+ (N+1)(n + n_f n_{\tilde{w}} + n_{\tilde{w}}^2n_D) + n_f^2)\end{aligned}$ \\
        \bottomrule 
        \end{tabular}
        \vspace*{0.15cm}
\end{table*}
\vspace{-0.15cm}
\section{SLTMPC with Asynchronous Updates}\label{sec:asynch-up}\vspace{-0.1cm}
Filter-based SLTMPC methods~\eqref{SLTMPC:generic},~\eqref{SLTMPC:rec-feas}, and~\cite{Chen2023} are computationally demanding due to the concurrent optimization of nominal trajectories~$\mathbf{z},\,\mathbf{v}, \, \mathbf{p}$, error dynamics~\eqref{eq:SLP-affine-constraint} parameterized by error system responses~$\bm{\Phi}^\mathbf{e},\, \bm{\Phi}^{\bm{\nu}}$ and disturbance filter~$\bm{\Sigma}, \, \bm{\Xi}$, and terminal set scaling~$\alpha$. The computational complexity of these methods is stated in Table~\ref{table:comp-complexity}. This complexity renders the methods unsuitable for practical applications with fast sampling times and limited computational resources. To alleviate some of the computational burden, we extend the asynchronous computation scheme introduced in~\cite{Sieber2023} for filter-based SLTMPC. To reduce the complexity of~\eqref{SLTMPC:rec-feas}, we remove error dynamics parameters~$\bm{\Phi}^\mathbf{e}, \,\bm{\Phi}^{\bm{\nu}}, \, \bm{\Sigma}, \,\bm{\Xi}$, and terminal scaling~$\alpha$ as optimization variables and instead use values that are computed at a lower frequency. The asynchronous computation scheme uses this idea to split the optimization of~\eqref{SLTMPC:rec-feas} into two separate processes - the primary and the secondary process. The secondary process computes the error dynamics parameters and the terminal set scaling, before passing them to the primary process, which only optimizes the nominal trajectories~$\mathbf{z},\,\mathbf{v}, \, \mathbf{p}$. The asynchronous computation scheme therefore enables running the primary process in closed-loop with system~\eqref{eq:dynamics} at a high frequency. The computationally expensive secondary process then runs concurrently at a lower frequency and reduces conservativeness by periodically providing new error dynamics parameters and terminal set scalings. Apart from limited computational resources, practical applications also impose memory constraints, i.e., we can only store a limited number of error dynamics parameters and terminal set scalings, which we handle by specific memory update rules. Figure~\ref{fig:asynch-up} provides an overview of this scheme.

\begin{figure}[h]
        \centering
        \input{tikz/asynchronous-updates}
        \caption{Visualization of the asynchronous computation scheme for filter-based SLTMPC: the secondary process computes error system responses~$\bm{\Phi}^\mathbf{e}\, \bm{\Phi}^{\bm{\nu}}$, disturbance filter~$\bm{\Sigma}, \, \bm{\Xi}$, and terminal set scaling~$\alpha$, which are then passed to the primary process and stored in memory~$\mathtt{M}$.}
        \label{fig:asynch-up} 
\end{figure}

\subsection{Secondary Process: Optimize Tubes}
The main objective of the secondary process is to compute error dynamics parameters~$\bm{\Phi}^\mathbf{e}$, $\bm{\Phi}^{\bm{\nu}}$, $\bm{\Sigma}, \, \bm{\Xi}$, and terminal set scaling~$\alpha$, which define the state, input, and disturbance tubes, and the terminal set. Therefore, we can choose the cost function of the secondary process according to any criterion the tubes and the terminal set should fulfill, e.g. a cost on~$\bm{\Phi}^\mathbf{e}$ promotes more aggressive tube controllers, thus leading to smaller error bounds on the state trajectories. For more details on possible cost functions and their effect on the tubes we refer to~\cite[Section~3.2]{Sieber2023}. We formulate the secondary process similarly to~\eqref{SLTMPC:rec-feas} but with a modified objective, i.e.,
\begin{subequations}\label{eq:process2}
\begin{align}
\min_{\bm{\Phi}^\mathbf{e}, \bm{\Phi}^{\bm{\nu}}, \bm{\Sigma}, \bm{\Xi}, \alpha \geq 0} \quad & L(\bm{\Phi}^\mathbf{e}, \bm{\Phi}^{\bm{\nu}}, \bm{\Sigma}, \bm{\Xi}), \\
\textrm{s.t. } \:\; & \eqref{SLTMPC-rec-feas:init} - \eqref{SLTMPC-rec-feas:terminal-dist-incl},
\end{align}
\end{subequations}
where $L(\cdot)$ is the chosen cost function. After termination of~\eqref{eq:process2}, we use the optimized error system responses~$\bm{\Phi}^\mathbf{e}$, $\bm{\Phi}^{\bm{\nu}}$ to compute the tightened state and input constraints for $i = 0,\mydots,N$ as
\begin{align}\label{process2-output:state-input-constr}
        \mathcal{Z}_i = \mathcal{X} \ominus \mathcal{F}_i(\bm{\Phi}^\mathbf{e}), \quad \mathcal{V}_i = \mathcal{U} \ominus \mathcal{F}_i(\bm{\Phi}^{\bm{\nu}}).
\end{align}
Additionally, we use the optimized disturbance filter~$\bm{\Sigma}, \, \bm{\Xi}$ to compute the disturbance tubes for~$d=1, \mydots, n_D$ and $i = 0,\mydots,N\!-\!1$ as
\begin{equation}
        \mathcal{Q}^d_{i} = \sigma_{i+1} \bar{\mathcal{W}} \ominus \mathcal{F}_i(\bm{\Psi}^d), \label{process2-output:dist-tube}
\end{equation}
where
\begin{equation*}
        \mathcal{F}_i(\bm{\Psi}^d) \coloneqq \mathcal{W} \oplus \bigoplus_{j=0}^{i-1} \Psi^d_{i,j} \bar{\mathcal{W}},
\end{equation*}
and $\Psi^d_{i,j}$ defined in~\eqref{eq:q_Pi}. For $i = N$, the disturbance tubes are computed as
\begin{equation}
        \mathcal{Q}^d_{N} = \sigma_{N} \bar{\mathcal{W}} \ominus \mathcal{F}_N(\bm{\Psi}^d). \label{process2-output:terminal-dist-tube}
\end{equation}
These sets are then passed to the primary process together with the error system responses, the disturbance filter, and the terminal set scaling as $\mathtt{M}_\textrm{new} = (\mathcal{Z}_{i},\, \mathcal{V}_{i}, \, \mathcal{Q}^d_{i}, \, \alpha, \,\bm{\Phi}^\mathbf{e}\!, \, \bm{\Phi}^{\bm{\nu}}\!, \, \bm{\Sigma}, \, \bm{\Xi})$. Since the memory $\mathtt{M} = \{ \mathtt{M}_{m} \mid m = 0,\mydots,M\scalebox{0.9}[1.0]{\( - \)}1 \}$ - where we denote the $m^\mathrm{th}$ memory entry as $\mathtt{M}_m$ - is finite, we need to define a procedure that updates the memory whenever the secondary process has computed a new~$\mathtt{M}_\mathrm{new}$, thus rendering the memory time-varying. In case the memory is full, the proposed update procedure in Algorithm~\ref{alg:memory-update-sec} selects a memory slot~$\mathtt{M}_{m}$ to store $\mathtt{M}_\textrm{new}$ according to a heuristic~$F(\mathtt{M}(k))$. For example,~$F(\mathtt{M}(k))$ can choose the slot with the least importance to the solution of the primary process or assign a score to each memory slot based on its historical importance and then choose the slot with the lowest score.
\begin{algorithm}[b]
        \caption{Update memory $\mathtt{M}$ (Secondary Process)}\label{alg:memory-update-sec}
        \hspace*{\algorithmicindent}\! \textbf{Input:} $\mathtt{M}_\mathrm{new}$, $\mathtt{M}(k\scalebox{0.9}[1.0]{\( - \)}1)$, $\bm{\lambda}(k\scalebox{0.9}[1.0]{\( - \)}1)$ \\
        \hspace*{\algorithmicindent}\! \textbf{Output:} $\mathtt{M}(k)$
        \begin{algorithmic}[1]
        \Procedure{updateMemory}{$\mathtt{M}_\mathrm{new}$}
        \State \(\triangleright\) runs when the secondary process terminates
        \State $\mathtt{M}(k) \gets \mathtt{M}(k\scalebox{0.9}[1.0]{\( - \)}1)$
        \If{$\mathtt{M}(k)$ not full}
                \State $m \gets$ index of first empty memory slot
                \State $\mathtt{M}_m(k) \gets \mathtt{M}_\mathrm{new}$
        \Else
                \State \(\triangleright\) determine a memory slot (except $m=0$)
                \State $m \gets F(\mathtt{M}(k))$
                \State $\mathtt{M}_m(k) \gets \mathtt{M}_\mathrm{new}$
        \EndIf
        \EndProcedure
        \end{algorithmic}
\end{algorithm}

\subsection{Primary Process: Optimize Nominal Trajectory}\vspace{-0.15cm}
Since a selection of tightened state and input constraints, disturbance tubes, and terminal sets is stored in memory~$\mathtt{M}$, the primary process can use this information to optimize the nominal trajectories. However, rather than using a single memory entry~$\mathtt{M}_m$, all entries are \emph{fused} using a convex combination similar to~\cite{Sieber2023,Kogel2020b}, ensuring all available information is optimally used. The primary process is formulated as\vspace{-0.12cm}
\begin{subequations}\label{eq:process1}
        \begin{align}
                \min_{\mathbf{z}, \mathbf{v}, \mathbf{p}, \bm{\lambda}} \quad & l_f(z_N) + \sum_{i=0}^{N-1} l(z_i, v_i), \label{process1:cost}\\ 
                \textrm{s.t. } \:\; & \forall\, i = 0, \mydots, N\!-\!1: \nonumber \\
                & z_0 = x(k), \\
                & z_{i+1} = Az_i + Bv_i + p_i, \label{process1:dynamics} \\
                & z_i \!\in\! \bigoplus_{m=0}^{M-1} \lambda_m\, \mathcal{Z}_i^m, \label{process1:state_constraints}\\
                & v_i \!\in\! \bigoplus_{m=0}^{M-1} \lambda_m \, \mathcal{V}_i^m, \\
                & z_N \!\in\! \bigoplus_{m=0}^{M-1} \lambda_m\, \alpha^m \,\mathcal{Z}_f, \label{process1:terminal-constraint} \\
                & \psi^d_i \!\in\! \bigoplus_{m=0}^{M-1} \lambda_m\, \mathcal{Q}^{d,m}_i, \quad \forall \:\! d=1,\mydots,n_D, \label{process1:dist-constraint}\\
                & \lambda_m \geq 0, \  \sum_{m=0}^{M-1} \lambda_m = 1, \label{process1:convex-comb}
                \vspace{-0.15cm}
        \end{align}
\end{subequations}
where $\bm{\lambda} = [ \lambda_0 \ \mydots \ \lambda_{M-1} ]$ are the convex combination variables, the costs $l_f(\cdot), \, l(\cdot,\cdot)$ fulfill Assumption~\ref{assump:cost}, $\psi^d_i$ is defined as in~\eqref{eq:q_Pi}, and~$\mathcal{Z}^m_i, \, \mathcal{V}^m_i, \,\mathcal{Q}^{d,m}_{i}, \, \alpha^m$ are retrieved from memory slot~$\mathtt{M}_m$. The resulting MPC control law is given by
\begin{equation}\label{eq:process1-law}
        \kappa_\textrm{primary}(x(k)) = v_0^*,
        \vspace{0.15cm}
\end{equation}
where $v_0^*$ is the first element of optimizer~$\mathbf{v}^*$ of~\eqref{eq:process1}.
Since the tubes computed by the secondary process are more general than those used in~\cite{Sieber2023}, we cannot use the proof in~\cite{Sieber2023} to show recursive feasibility of the primary process. Therefore, we additionally store the previous solution of~\eqref{eq:process1} in memory slot~$\mathtt{M}_0$ as a fallback solution. Specifically, we store the shifted convex combination of tubes after the primary process terminates, i.e., for all $d=1, \mydots, n_D$ and $i = 0, \mydots, N\!-\!1$ we compute \vspace{-0.15cm}
\begin{subequations}\label{process1:shifted-solution}
\begin{align}
        \mathcal{Z}_i^\mathrm{prev} &= \!\bigoplus_{m=0}^{M-1} \lambda_m^* \left(\mathcal{X} \ominus \mathcal{F}_i (\bm{\Phi}^{\mathbf{e}, m}_\mathrm{shift})\right), \\
        \mathcal{V}_i^\mathrm{prev} &= \!\bigoplus_{m=0}^{M-1} \lambda_m^* \left(\mathcal{U}\ominus \mathcal{F}_i (\bm{\Phi}^{\bm{\nu}, m}_\mathrm{shift})\right), \\
        \mathcal{Q}^{d, \mathrm{prev}}_i &= \!\bigoplus_{m=0}^{M-1} \lambda_m^* \left( \sigma_{i+2}^m \bar{\mathcal{W}} \ominus \mathcal{F}_i (\bm{\Psi}^{d, m}_\mathrm{shift}) \right),
        \vspace{-0.1cm}
\end{align}
\end{subequations}
where the shifted system responses are computed as
\begin{subequations}\label{eq:shifted-system-responses}
\begin{align}
        \bm{\Phi}^{\mathbf{e}, m}_\mathrm{shift} &= \mathrm{shift}(\bm{\Phi}^{\mathbf{e},m}, \bm{\Phi}^{\mathbf{e},m}_{N,:}), \\
        \bm{\Phi}^{\bm{\nu}, m}_\mathrm{shift} &= \mathrm{shift}(\bm{\Phi}^{\bm{\nu},m}, \bm{\Phi}^{\bm{\nu},m}_{N,:}),
\end{align}     
\end{subequations}
with the $\mathrm{shift}(\cdot,\cdot)$ operator defined in Section~\ref{sec:notation} and $\bm{\Phi}_{N,:}$ denoting the last block row of~$\bm{\Phi}^{\mathbf{e},m}$ and $\bm{\Phi}^{\bm{\nu},m}$. The shifted auxiliary variable~$\bm{\Psi}^{d, m}_\mathrm{shift}$ is computed with~\eqref{eq:q_Pi} using the shifted system responses and the shifted disturbance filter, i.e.,
\begin{subequations}\label{eq:shifted-dist-filter}
\begin{align}
        \bm{\Sigma}^{m}_\mathrm{shift} &= \mathrm{shift}(\bm{\Sigma}^m, [\bm{\Xi}^m_{1:} \ \sigma_N^m \cdot \mathbb{I}_n]), \\
        \bm{\Xi}^{m}_\mathrm{shift} &=  \bm{\Xi}^m,
\end{align}
\end{subequations}
where $\bm{\Xi}^m_{1:} = [ \Xi_1^m \ \mydots \ \Xi_{N-1}^m]$ and the $\sigma_i$ are obtained from either~$\bm{\Phi}^{\mathbf{e}, m}$ or~$\bm{\Sigma}^m$ using~\eqref{proof:SLP-dyn-init}. Additionally, we compute the tubes for $i=N$ as
\begin{align*}
        \mathcal{Z}_N^\mathrm{prev} &= \!\bigoplus_{m=0}^{M-1} \lambda_m^* \, \mathcal{Z}_{N}^m, \quad
        \mathcal{V}_N^\mathrm{prev} = \!\bigoplus_{m=0}^{M-1} \lambda_m^* \, \mathcal{V}_{N}^m, \\
        \mathcal{Q}^{d, \mathrm{prev}}_N &= \!\bigoplus_{m=0}^{M-1} \lambda_m^* \, \mathcal{Q}^{d, m}_N.
\end{align*}
Apart from the tubes, we also store the fused terminal set scaling, the shifted error system responses, and the shifted disturbance filter, since they are needed to recompute~\eqref{process1:shifted-solution} in case $\lambda^*_0 \neq 0$ in the next solution of~\eqref{eq:process1}. Therefore, we compute $\alpha^\mathrm{prev} = \sum_{m=0}^{M-1} \lambda_m^* \alpha^m$ and
\begin{alignat*}{2}
        \bm{\Phi}^{\mathbf{e}, \mathrm{prev}} &= \!\sum_{m=0}^{M-1} \lambda_m^* \, \bm{\Phi}^{\mathbf{e}, m}_\mathrm{shift}, \quad
        \bm{\Phi}^{\bm{\nu},\mathrm{prev}} &&= \!\sum_{m=0}^{M-1} \lambda_m^* \, \bm{\Phi}^{\bm{\nu}, m}_\mathrm{shift},\\
        \bm{\Sigma}^\mathrm{prev} &= \!\sum_{m=0}^{M-1} \lambda_m^*\, \bm{\Sigma}^{m}_\mathrm{shift},\quad
        \bm{\Xi}^\mathrm{prev} &&= \!\sum_{m=0}^{M-1} \lambda_m^*\,\bm{\Xi}^{m}_\mathrm{shift},
\end{alignat*}
before collecting all computed sets and parameters in memory entry
\begin{align}\label{eq:memory-update-proc1}
\mathtt{M}_\textrm{prev} = (&\mathcal{Z}_i^\mathrm{prev}, \, \mathcal{V}_i^\mathrm{prev},\, \mathcal{Q}_i^{d,\mathrm{prev}}\!,\, \alpha^\mathrm{prev}, \nonumber\\
        &\bm{\Phi}^{\mathbf{e},\mathrm{prev}}\!,\, \bm{\Phi}^{\bm{\nu},\mathrm{prev}}\!,\, \bm{\Sigma}^\mathrm{prev}\!,\, \bm{\Xi}^\mathrm{prev}),
\end{align}
and updating the memory according to Algorithm~\ref{alg:memory-update-pri}.
\begin{remark}
        In practice, the shifting operations~\eqref{eq:shifted-system-responses},~\eqref{eq:shifted-dist-filter} can be performed by the secondary process after computing new~$\bm{\Phi}^\mathbf{e}$, $\bm{\Phi}^{\bm{\nu}}$, $\bm{\Sigma}, \, \bm{\Xi}$. Due to the specific shifting operation, these computations only need to be performed $N$ times. Therefore, also the tightened sets in~\eqref{process1:shifted-solution}, e.g. $\mathcal{X} \ominus \,\mathcal{F}_i (\bm{\Phi}^{\mathbf{e}, m}_\mathrm{shift})$, can be computed by the secondary process. The primary process then just performs the convex combination in~\eqref{process1:shifted-solution}, which only marginally increases the computational complexity of the primary process.
\end{remark}
\begin{algorithm}[t]
\caption{Update memory $\mathtt{M}$ (Primary Process)}\label{alg:memory-update-pri}
\hspace*{\algorithmicindent}\! \textbf{Input:} $\mathtt{M}_\mathrm{prev}$, $\mathtt{M}(k\scalebox{0.9}[1.0]{\( - \)}1)$ \\
\hspace*{\algorithmicindent}\! \textbf{Output:} $\mathtt{M}(k)$
\begin{algorithmic}[1]
\Procedure{updateMemory}{$\mathtt{M}_\mathrm{prev}$}
\State \(\triangleright\) runs when the primary process terminates
\State $\mathtt{M}(k) \gets \mathtt{M}(k\scalebox{0.9}[1.0]{\( - \)}1)$
\State $\mathtt{M}_0(k) \gets \mathtt{M}_\mathrm{prev}$ where $\mathtt{M}_\mathrm{prev}$ is given by~\eqref{eq:memory-update-proc1}
\EndProcedure
\end{algorithmic}
\end{algorithm}

\subsection{Theoretical Guarantees}\vspace{-0.2cm}
In the following, we provide theoretical guarantees for the proposed asynchronous computation scheme applied to~\eqref{eq:dynamics}, i.e., we prove recursive feasibility and ISS of primary process~\eqref{eq:process1} for any state of the memory, including the case in which the memory is updated. For the proofs, we make use of the following assumption.\vspace{-0.1cm}
\begin{assumption}\label{assump:init-feas}
At the start of the control task, memory~$\mathtt{M}$ contains at least one memory entry~$\mathtt{M}_0$ for which the primary process is feasible, i.e.,~\eqref{eq:process1} is feasible for $\mathtt{M}_0 = (\mathcal{Z}_i^0, \, \mathcal{V}_i^0,\, \mathcal{Q}_i^{d,0},\, \alpha^0, \, \bm{\Phi}^{\mathbf{e},0},\, \bm{\Phi}^{\bm{\nu}, 0},\, \bm{\Sigma}^0,\, \bm{\Xi}^0)$ with $\lambda_0 = 1$.
\end{assumption}
\vspace{-0.1cm}
Note that Assumption~\ref{assump:init-feas} is satisfied, if for example the secondary process is once run offline and the memory is initialized with the computed memory entry~$\mathtt{M}_\mathrm{new}$.\vspace{-0.1cm}
\begin{prop}\label{prop:rec-feas-proc1}
Let Assumption~\ref{assump:init-feas} hold. Then, primary process~\eqref{eq:process1} is recursively feasible for any state of memory~$\mathtt{M}(k)$ and any memory updates according to Algorithms~\ref{alg:memory-update-sec} and~\ref{alg:memory-update-pri}.
\end{prop}
\vspace{-0.4cm}
\begin{pf}
We prove the proposition using the standard shifting argument~\cite{Rawlings2009} and relying on the proof of Theorem~\ref{prop:rec-feas}. We construct the candidate sequences specifically for the case in which we only use memory slot~$\mathtt{M}_0$ that contains the shifted previous tubes~\eqref{eq:memory-update-proc1}. This allows us to show recursive feasibility for any state of the memory, including the case in which the memory is updated, since we can always fallback to the tubes in~$\mathtt{M}_0$. Due to Assumption~\ref{assump:init-feas}, the primary process is then guaranteed to be recursively feasible.
\newline
Let $\mathbf{z}^* = [ z_0^{* \top}, \mydots, z_N^{* \top} ]^\top$, $\mathbf{v}^* = [v_0^{* \top}, \mydots, v_{N-1}^{* \top}]^\top$, $\mathbf{p}^* = [p_0^{* \top}, \mydots, p_{N-1}^{* \top}]^\top$, and $\bm{\lambda}^* = [\lambda_0^*, \mydots, \lambda_{M-1}^* ]^\top$ denote the optimal solution of~\eqref{eq:process1} for initial state $x(k)$. We then use the candidate $\hat{\lambda}_0 = 1$, $\hat{\lambda}_m = 0, \, m=1,\mydots,M\scalebox{0.75}[1.0]{\( - \)}1$ for the convex combination variables, which trivially fulfills constraint~\eqref{process1:convex-comb}, and construct the remaining candidate sequences as\vspace{-0.1cm}
\begin{subequations}\label{proof:candidates}
\begin{alignat}{2}
\hat{z}_i &\!=\! z_{i+1}^* \!+\! \!\sum_{m=0}^{M-1} \!\lambda_m^*\, \Phi^{e, m}_{i+1,0} \bar{w}(k), &&\ i\!=\!0,\mydots,N\!-\!1,\\
\hat{z}_N &\!=\! A_{K_f}z_N^* \!+\! \!\sum_{m=0}^{M-1} \!\lambda_m^*\, \Gamma^m \bar{w}(k), \\
\hat{v}_i &\!=\! v_{i+1}^* \!+\! \!\sum_{m=0}^{M-1} \!\lambda_m^*\, \Phi^{\nu,m}_{i+1,0} \bar{w}(k), &&\ i\!=\!0,\mydots,N\!-\!2,
\vspace{-0.1cm}
\end{alignat}
\begin{alignat}{2}
\hat{v}_{N\scalebox{0.5}[1.0]{\( - \)}1} &\!=\! K_f z_{N}^* \!+\! \!\sum_{m=0}^{M-1} \!\lambda_m^*\Phi^{\nu,m}_{N,0} \bar{w}(k), \\
\hat{p}_i &\!=\! p_{i+1}^* \!+\! \!\sum_{m=0}^{M-1} \!\lambda_m^*\, \Sigma^m_{i+2,0} \bar{w}(k), &&\ i\!=\!0,\mydots,N\!-\!2, \\
\hat{p}_{N\scalebox{0.5}[1.0]{\( - \)}1} &\!=\! \sum_{m=0}^{M-1} \!\lambda_m^*\,\Xi^m_{0} \bar{w}(k),
\vspace{-0.15cm}
\end{alignat}
\end{subequations}
with $\Gamma^m = A \Phi^{e, m}_{N,0} + B \Phi^{\nu,m}_{N,0} + \Xi^m_0$. Next, we show that candidates~\eqref{proof:candidates} satisfy all constraints using the proof of Theorem~\ref{prop:rec-feas}. Note that~\eqref{proof:candidates} are equivalent to the vector-valued candidates in~\eqref{proof:vec-cand} and the tubes stored in~$\mathtt{M}_0$ are constructed equivalently to~\eqref{proof:mat-cand}, apart from the convex combination of the system responses and disturbance filters. However, since the candidates and the tubes use the same convex combination, we can treat each contribution to the convex combination individually. For example, to show $\hat{z}_i \in \mathcal{Z}_i^0, \, i = 0, \mydots, N\!-\!1$, we get\vspace{-0.15cm}
\begin{align*}
\sum_{m=0}^{M-1}\!\lambda_m^*\, (z_{i+1}^* \!+ \Phi^{e, m}_{i+1,0}\, \bar{w}(k)) \in \!\bigoplus_{m=0}^{M-1} \!\lambda_m^* \left(\mathcal{X} \ominus \mathcal{F}_i (\bm{\Phi}^{\mathbf{e}, m}_\mathrm{shift})\right),
\vspace{-0.15cm}
\end{align*}
and we can show $z_{i+1}^* + \Phi^{e, m}_{i+1,0}\, \bar{w}(k) \in  \mathcal{X} \ominus \mathcal{F}_i(\bm{\Phi}^{\mathbf{e}, m}_\mathrm{shift})$ for each $m$ individually. As shown in the proof of Theorem~\ref{prop:rec-feas}, tubes~\eqref{process1:shifted-solution} satisfy this inclusion by construction. The same argument holds for the other constraints and thus we have shown that candidate~\eqref{proof:candidates} satisfies all constraints in~\eqref{eq:process1}, which proves recursive feasibility.\qed
\end{pf}
\vspace{-0.35cm}
For robust stability, we note that costs~$l(\cdot), \, l_f(\cdot)$ in~\eqref{process1:cost} are only functions of the nominal trajectories and independent of the convex combination variables. Therefore, the following stability result is a direct consequence of Theorem~\ref{prop:ISS} and Proposition~\ref{prop:rec-feas-proc1}.\vspace{-0.15cm}
\begin{cor}\label{prop:ISS-proc1}
Given that Assumption~\ref{assump:cost} holds, system~\eqref{eq:dynamics} subject to constraints~\eqref{eq:constraints} and in closed-loop with~\eqref{eq:process1-law}, is ISS in $\mathcal{X}_\textrm{feas}$ for any admissible sequence of combined disturbances $\bm{\eta}$, where $\mathcal{X}_\textup{feas}$ is the set of all states $x(k)$ for which~\eqref{eq:process1} is feasible.
\end{cor}
\vspace{-0.15cm}
\begin{remark}
        In case heuristic $F(\mathtt{M}(k))$ enters the cost of~\eqref{eq:process1}, e.g. as a regularizer, ISS as in Corollary~\ref{prop:ISS-proc1} cannot be proven. However, we can prove input-to-state practical stability (ISpS)~\cite{Limon2009}, since the heuristic term is constant with respect to the state and disturbance.
\end{remark}
\vspace{-0.1cm}
\begin{figure*}[t]
        \centering
        \includegraphics[width=1\linewidth]{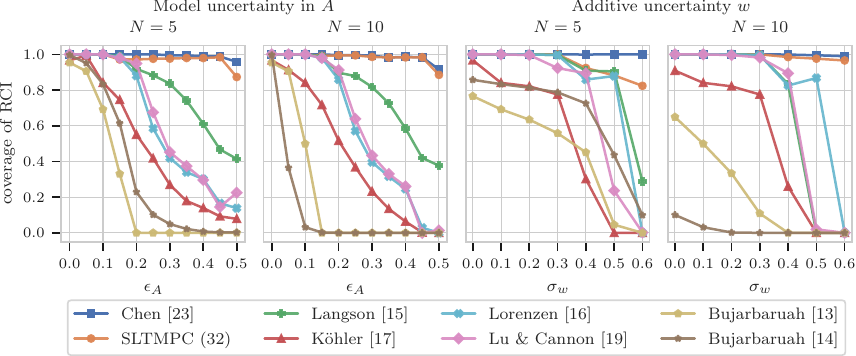}
        \vspace{-0.17cm}
        \caption{Comparison of~\eqref{SLTMPC:rec-feas} to~\cite{Chen2023} and other standard tube-based MPC methods~\cite{Bujarbaruah2021,Bujarbaruah2022,Langson2004,Lorenzen2019,Kohler2019,Lu2019}: Region of attraction (RoA) relative to the maximal RCI set in percent, where we vary the uncertainty in the dynamics matrix~$A$, i.e. $\epsilon_A$, (left) and the additive disturbance, i.e. $\sigma_w$, (right). The other uncertainty parameters are fixed to $\epsilon_B = 0.1$, $\sigma_w = 0.1$ and $\epsilon_A = 0.1$, $\epsilon_B = 0.1$, respectively, and the horizon is $N=\{5, \, 10\}$.}
        \label{fig:Chen_comparison}
\end{figure*}
\vspace{-0.1cm}
\section{Numerical Results}\label{sec:numerical_section}\vspace{-0.2cm}
We first compare recursively feasible SLTMPC~\eqref{SLTMPC:rec-feas} to the method presented in~\cite{Chen2023} and standard tube-based MPC methods from the literature~\cite{Bujarbaruah2021,Bujarbaruah2022,Langson2004,Lorenzen2019,Kohler2019,Lu2019} on a two dimensional double integrator example, before showing the benefits of the asynchronous computation scheme on a more complex example. All examples are implemented in Python using CVXPY~\cite{cvxpy} and are solved using MOSEK~\cite{mosek}. The examples were run on a machine equipped with a~\unit[3.1]{GHz} CPU and \unit[32]{GB} of RAM.

\subsection{Double Integrator}
In the first example, we use the same system as in~\cite{Chen2023}, i.e., uncertain LTI system~\eqref{eq:dynamics} with discrete-time dynamic matrices
\begin{equation*}
A = \begin{bmatrix} 1 & 0.15 \\ 0.1 & 1 \end{bmatrix}\!, \quad B = \begin{bmatrix} 0.1 \\ 1.1 \end{bmatrix}\!,
\end{equation*}
subject to polytopic constraints $\| x \|_\infty \leq 8$, $\| u \|_\infty \leq 4$, additive disturbance $\| w \|_\infty \!\leq\! \sigma_w$, and model uncertainty
\begin{equation*}
\Delta_A \!\in\! \textrm{co}\left\lbrace \begin{bmatrix} \epsilon_A & 0 \\ 0 & 0 \end{bmatrix}\!, \begin{bmatrix} \scalebox{0.7}[1.0]{\( - \)}\epsilon_A & 0 \\ 0 & 0 \end{bmatrix}\right\rbrace\!, \Delta_B \!\in\! \textrm{co}\left\lbrace \begin{bmatrix} 0 \\ \epsilon_B \end{bmatrix}\!, \begin{bmatrix} 0 \\ \scalebox{0.7}[1.0]{\( - \)}\epsilon_B \end{bmatrix}\right\rbrace\!,
\end{equation*}
where the uncertainty parameters $\sigma_w, \, \epsilon_A$, and $\epsilon_B$ are varied. Additionally, we use the cost function~$l(z,v) = z^\top\! Qz + v^\top\! Rv$ with $Q=10\cdot \mathbb{I}_{2}$ and $R=1$, and horizon $N \in \{5, 10\}$. As the terminal sets, we use the maximal robust control invariant (RCI) set of~\eqref{eq:dynamics} for~\cite{Chen2023} and the maximal RPI set (Definition~\ref{def:RPI}) - with $K_f$ the LQR controller - for~\eqref{SLTMPC:rec-feas}. We design the auxiliary disturbance set $\bar{\mathcal{W}}$ to be equivalent to $\mathcal{W}$ for both methods and since these are hyperrectangles we use Remark~\ref{remark:SLTMPC-hyperrectangle} to modify~\eqref{SLTMPC:rec-feas}. The setup of the other methods is equivalent to~\cite[Section~6]{Chen2023} and we refer to~\cite{Chen2023} for further details.
\begin{table*}
        \centering
        \caption{Average computation times for the RoA computations in the double integrator example (Figure~\ref{fig:Chen_comparison}).}\label{table:comp-times-RoA}
        \vspace*{0.12cm}
        \begin{tabular}{@{}lccccccccc@{}}
        \toprule
        & $N$ & ours~\eqref{SLTMPC:rec-feas} & Chen~\cite{Chen2023} & Langson~\cite{Langson2004} & Köhler~\cite{Kohler2019} & Lorenzen~\cite{Lorenzen2019} & Lu~\cite{Lu2019} & Bujarbaruah~\cite{Bujarbaruah2021} $\mid$ \cite{Bujarbaruah2022}\\ \midrule
        \multirow{2}{*}{Time [ms]} & 5 & 33.9 & 27.4 & 53.5 & 5.3 & 644.2 & 7.9 & \!11.2 $\mid$ 28.3\\
        & 10 & 99.5 & 83.1 & 107.6 & 6.36 & 1958.8 & 10.0 & \ 47.6 $\mid$ 104.9\\
        \bottomrule 
        \end{tabular}
\end{table*}

Figure~\ref{fig:Chen_comparison} shows the region of attraction~(RoA), i.e., the set of initial states for which the MPC method is feasible, for all considered methods as a fraction of the maximal RCI set for dynamics~\eqref{eq:dynamics}. We vary the parametric uncertainty~$\epsilon_A \in [0.0, 0.5]$, while fixing~$\epsilon_B = \sigma_w = 0.1$ (left) and vary the additive disturbance~$\sigma_w \in [0.0, 0.6]$, while fixing~$\epsilon_A = \epsilon_B = 0.1$ (right). We notice that both the method in~\cite{Chen2023} and \eqref{SLTMPC:rec-feas} outperform the other methods, especially for larger uncertainties. Generally, the method in~\cite{Chen2023} is less conservative than~\eqref{SLTMPC:rec-feas}, i.e., its RoA covers a larger part of the maximal RCI set. This is due to the additional constraints in~\eqref{SLTMPC:rec-feas} needed for recursive feasibility and the use of an RCI terminal set compared to the RPI terminal set in~\eqref{SLTMPC:rec-feas}. However, with increasing horizon~$N$ the RoAs of the two methods become more similar. This is because the method in~\cite{Chen2023} tightens the RCI terminal set with the disturbance overapproximation, which becomes more conservative for larger horizons, thus reducing the RoA. In contrast, the RoA of~\eqref{SLTMPC:rec-feas} increases with the horizon, since the MPC controller is enhancing the terminal RPI set, thus reducing conservativeness with larger horizons. The average computation times for the RoA computations are stated in Table \ref{table:comp-times-RoA}.
\begin{figure}[t]
        \centering
        \includegraphics[width=0.99\columnwidth]{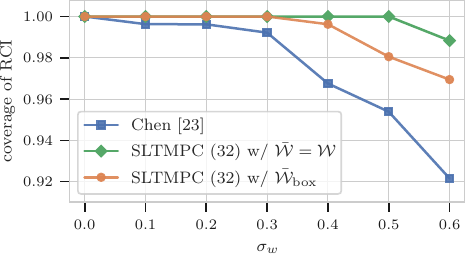}
        \vspace{-0.1cm}
        \caption{RoA relative to the maximal RCI set in percent for \eqref{SLTMPC:rec-feas} and \cite{Chen2023}, where we vary $\sigma_w$ and the choice of $\bar{\mathcal{W}}$, but keep $\epsilon_A = \epsilon_B = 0.1$ and $N=5$ fixed.}
        \label{fig:W_tilde}
\end{figure}

\begin{figure}[t]
        \centering
        \includegraphics[width=0.98\columnwidth]{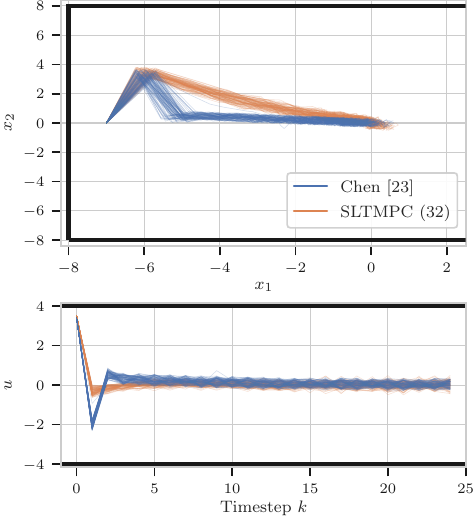}
        \vspace{-0.1cm}
        \caption{Closed-loop trajectories of length $25$ for~\cite{Chen2023} and~\eqref{SLTMPC:rec-feas}. The average costs are $2143.7$ for~\cite{Chen2023} and $1951.1$ for~\eqref{SLTMPC:rec-feas}.}
        \label{fig:perf_comparison}
\end{figure}

To analyze the choice of $\bar{\mathcal{W}}$ on the performance, we compare the method in~\cite{Chen2023} to \eqref{SLTMPC:rec-feas} for the same setup as in Figure~\ref{fig:Chen_comparison}~(right) but with a different uncertainty description $\mathcal{W}$, i.e.~\eqref{eq:W} with $h_w = (0.5\,\sigma_w,\, \sigma_w,\, 0.5\,\sigma_w,\, 0.5\,\sigma_w,\, \sigma_w,\, 0.5\,\sigma_w)$. While $\bar{\mathcal{W}}$ is fixed for~\cite{Chen2023}, we choose $\bar{\mathcal{W}}$ once as a hyperrectangle~$\bar{\mathcal{W}}_\textrm{box}$ and once as $\bar{\mathcal{W}} = \mathcal{W}$ for \eqref{SLTMPC:rec-feas}. The results are shown in Figure \ref{fig:W_tilde}, where we note that \eqref{SLTMPC:rec-feas} is less conservative than the method in~\cite{Chen2023}, since the chosen $\bar{\mathcal{W}}$ enable better overapproximations of the system uncertainties $\mathcal{W},\, \mathcal{D}$. Additionally, the method in~\cite{Chen2023} has significant limitations, i.e., it requires a shrinking horizon implementation and a RPI set for dynamics~\eqref{eq:dynamics}, both of which are highlighted in the following. First, we compare the closed-loop performance of the two methods. Figure~\ref{fig:perf_comparison} shows $200$ closed-loop state and input trajectories starting in $x(0) = [-7, 0]^\top$. The average cost of the closed-loop trajectories are $2143.7$ for~\cite{Chen2023} and $1951.1$ for~\eqref{SLTMPC:rec-feas}, which is a $9\%$ improvement. This behavior is expected since~\eqref{SLTMPC:rec-feas} is applied in receding horizon, while the method in~\cite{Chen2023} is applied for a shrinking horizon. This is most evident in the first $5$ timesteps, where~\cite{Chen2023} exhibits abrupt changes in both the state and input trajectories.

\subsection{Vertical Take-off and Landing (VTOL) vehicle}
Next, we show-case the proposed filter-based SLTMPC on a VTOL vehicle model. We use~\cite{Bouffard2012} to derive the dynamics of a VTOL vehicle - inspired by the RockETH~\cite{Spannagl2021} - moving in a two-dimensional plane as depicted in Figure~\ref{fig:rocket}. Then, we obtain the discrete-time dynamics
\begin{subequations}\label{eq:VTOL-dynamics}
\begin{alignat}{2}
p^x_{i+1} &\!= p^x_{i} \!+\! \Delta_tv^x_{i}, \quad && p^z_{i+1} \!= p^z_{i} \!+\! \Delta_tv^z_{i},\\
v^x_{i+1} &\!= v^x_{i} \!+\! \Delta_t k_1 \theta_{i}, \quad && v^z_{i+1} \!= v^z_{i} \!+\! \Delta_t u^z_{i}, \\
\theta_{i+1} &\!= \theta_{i} \!+\! \Delta_t \omega_{i}, \quad && \omega_{i+1} \!= \Delta_t k_2 \theta_{i} \!+\! \omega_{i} \!+\! \frac{\Delta_t}{I} u^\theta_{i},
\end{alignat}    
\end{subequations}
where $p^{x/z}$ and $v^{x/z}$ denote the position and velocity of the VTOL vehicle in $x$ and $z$ direction, respectively, $\theta$ and $\omega$ are the angle and angular velocity of the vehicle with respect to the upright position, $u^z$, $u^\theta$ denote the control inputs, the parameters $k_1$, $k_2$ define the interaction between the translational and rotational dynamics, $I$ is the inertia, and $\Delta_t$ is the discretization time. We choose all parameters such that they mimic the RockETH~\cite{Spannagl2021} and constrain its position to $[\unit[-15]{m}, \unit[15]{m}]$ and $[\unit[0]{m}, \unit[15]{m}]$ in the $x$ and $z$ direction, respectively. The control inputs are constrained to $[\unit[-5]{N}, \unit[5]{N}]$ and $[\unit[-0.5]{Nm}, \unit[0.5]{Nm}]$ for $u^z$ and $u^\theta$, respectively. For further details on the experimental setup and the reformulation as system~\ref{eq:dynamics}, see Appendix~\ref{app:rocket-details}.
\begin{figure}[t]
        \centering
        \input{tikz/rocket}
        \vspace{-0.05cm}
        \caption{Visualization of the experimental setup for the VTOL vehicle.}
        \label{fig:rocket}
        \vspace{-0.15cm}
\end{figure}

\begin{figure}[t]
        \centering
        \includegraphics[width=0.98\columnwidth]{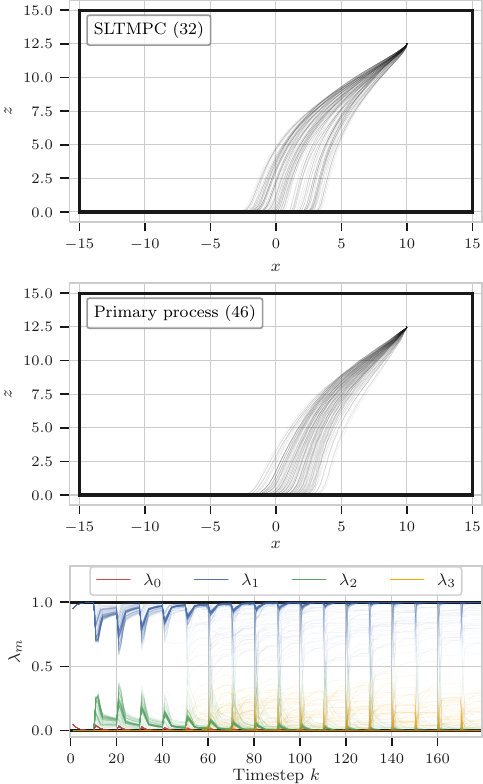}
        \vspace{-0.1cm}
        \caption{Closed-loop trajectories in the $x\scalebox{0.7}[1.0]{\( - \)}z$ plane of the VTOL vehicle with SLTMPC~\eqref{SLTMPC:rec-feas} (top) and the primary process~\eqref{eq:process1} (middle), and the corresponding convex combination parameters~$\lambda_m,\, m=0, \mydots, M\!-\!1$ (bottom).}
        \label{fig:rocket-experiment}
        \vspace{0.1cm}
\end{figure}

The interaction parameters $k_1, \, k_2$ are uncertain and we assume a $\pm 5\%$ modelling error on each of these parameters, which we model as the parametric uncertainty $\Delta_A$. Additionally, we assume wind gusts of $\pm \unit[2]{m/s}$ in $x$~direction, which we model as the additive disturbance~$w$. We assume no uncertainty in the input matrix~$B$ nor any uncertainty affecting the $z$~direction. For the given uncertainties, it is not possible to compute a polytopic RPI set for dynamics~\eqref{eq:dynamics}, therefore rendering both~\eqref{SLTMPC:generic} and the method in~\cite{Chen2023} inapplicable. However, we can compute an RPI set for auxiliary dynamics~\eqref{eq:auxiliary-dynamics} with only the additive disturbance set~$\bar{\mathcal{W}} = \{ \bar{w} \in \mathbb{R}^6 \mid \| \bar{w} \|_\infty \leq 0.075 \}$. In the following, we apply~\eqref{SLTMPC:rec-feas} and the asynchronous computation scheme~\eqref{eq:process1},~\eqref{eq:process2} with memory size $M=4$ to the VTOL vehicle for a horizon of~$N=10$. We show the closed-loop trajectories in the $x\scalebox{0.7}[1.0]{\( - \)}z$ plane for~\eqref{SLTMPC:rec-feas} and the primary process~\eqref{eq:process1} in Figure~\ref{fig:rocket-experiment} and note that both~\eqref{SLTMPC:rec-feas} and the primary process with asynchronously computed tubes~\eqref{eq:process1} are able to steer the vehicle to the origin. Due to the secondary process~\eqref{eq:process2} only updating the tubes every $10^\textrm{\:\!th}$ timestep, the SLTMPC with asynchronous computation has a higher average closed-loop cost of $5163.6$ compared to $4799.4$ for~\eqref{SLTMPC:rec-feas} However, the asynchronous computation is significantly faster than~\eqref{SLTMPC:rec-feas}, as shown in Table~\ref{table:comp-times-rocket}.

Figure~\ref{fig:rocket-experiment} also shows the convex combination parameters~$\lambda_m$ for the closed-loop simulations. The memory slots~$\mathtt{M}_0$ and~$\mathtt{M}_1$ are both initialized with the tubes computed by~\eqref{eq:process2} for initial condition~$[x, z]^\top = [10, 12.5]^\top$\!, memory slot~$\mathtt{M}_3$ is initialized with tubes computed by~\eqref{eq:process2} for initial condition~$[x, z]^\top = [0, 0]^\top$\!, and memory slot~$\mathtt{M}_2$ is left empty. The memory is then updated by Algorithm~\ref{alg:memory-update-pri} in every timestep and by Algorithm~\ref{alg:memory-update-sec} every ten timesteps, where the new tubes are stored in~$\mathtt{M}_1$ and the previous ones are copied to~$\mathtt{M}_2$. In Figure~\ref{fig:rocket-experiment}, these updates are clearly visible by the spikes in the convex combination variables. The spikes occur because we copy the previous tubes to~$\mathtt{M}_2$, i.e., when the tubes in~$\mathtt{M}_1$ (blue lines) are updated, it is beneficial to fuse them with the previous ones in~$\mathtt{M}_2$ (green lines). However, after a few timesteps it is preferable to mainly use the new tubes, which is evident by increasing~$\lambda_1$ (blue) and decreasing~$\lambda_2$ (green) towards the next tube update. For $k \geq 50$, the trajectories are approaching the origin and the tubes in~$\mathtt{M}_3$ (yellow lines) are increasingly used, which is expected since they are optimized for the region around the origin. Finally, we discourage the use of the fallback tubes in~$\mathtt{M}_0$ (red lines) with a regularization cost on $\lambda_0$, which is evident by their low usage, i.e., these tubes are only used if necessary to retain feasibility.

\section{Conclusions}\label{sec:conclusions}
This paper proposed filter-based system level tube-MPC (SLTMPC) for constrained discrete-time linear systems with additive disturbances and model uncertainties. We introduced a new terminal controller design combined with an online optimized terminal set, which allowed us to show rigorous closed-loop guarantees in receding horizon for the first time for this type of MPC method. Additionally, we introduced a new asynchronous computation scheme for filter-based SLTMPC, which significantly reduces its computational demand by separating the tube and nominal trajectory optimizations into different processes. Finally, we showed the benefits and effectiveness of the proposed methods on two numerical examples. In future work, the proposed method can be extended to other uncertainty descriptions and potentially online-updated uncertainty sets.
\begin{table}[t]
        \centering
        \caption{Computation times for the VTOL vehicle experiment with horizon ${N=10}$.}\label{table:comp-times-rocket}
        \vspace*{0.15cm}
        \begin{tabular}{@{}lcc@{}}
        \toprule
        & \multicolumn{2}{c}{Comp. Time [ms]} \\ \cmidrule(lr){2-3}
        & Min. & Median \\ \midrule
        SLTMPC~\eqref{SLTMPC:rec-feas} & 490.75 & 719.76 \\
        \midrule
        Asynchronous computation & & \\ \cmidrule(lr){1-1}
        Primary Process~\eqref{eq:process1} & 7.79 & 16.59 \\
        Secondary Process~\eqref{eq:process2} & 473.5 & 685.3 \\
        \bottomrule 
        \end{tabular}
\end{table}

\newpage
\bibliographystyle{ieeetr}
\bibliography{bibliography}


\appendix
\section{Implementation Details}
In this appendix, we discuss how the inclusion constraints in filter-based SLTMPC~\eqref{SLTMPC:rec-feas} and in the primary process~\eqref{eq:process1} can be implemented as linear constraints. To start, we formally define support functions~\cite{Kolmanovsky1998} below.
\begin{definition}[Support function]\label{def:sup_fn}
The support function of a non-empty compact convex set $\mathcal{A} \subset \mathbb{R}^n$, evaluated at $\xi \in \mathbb{R}^n$ is given by $h_\mathcal{A}(\xi) = \sup_{a \in \mathcal{A}} \xi^\top a$.
\end{definition}
The support function of a polytopic set $\mathcal{A} = \{ a \mid H_A a \leq h_a\}$ is then given by the solution to the linear program $\max_{a \in \mathcal{A}} \xi^\top a$.

\subsection{Implementation of filter-based SLTMPC~\eqref{SLTMPC:rec-feas}}\label{app:implementation}
While the state and input constraints~\eqref{SLTMPC-rec-feas:state-constraint},~\eqref{SLTMPC-rec-feas:input-constraint} can be reformulated as linear constraints using standard techniques from the disturbance feedback literature~\cite{Goulart2006} -- for a detailed discussion we refer to~\cite[Appendix~3.B]{phdthesis}, -- the other inclusion constraints require more consideration. The terminal constraints~\eqref{SLTMPC-rec-feas:terminal-constraint} -~\eqref{SLTMPC-rec-feas:terminal-decrease} are all special cases of the following set inclusion
\begin{equation*}
        \alpha A \mathcal{X} \subseteq \beta \mathcal{Y} \ominus \Gamma \mathcal{Z},
\end{equation*}
where $\alpha,\, \beta$ are scalars, $A, \, \Gamma$ are matrices, and $\mathcal{X}, \, \mathcal{Y}$ are polytopic sets, and $\mathcal{Z}$ is a compact convex set. Therefore, we can use the following lemma from~\cite{Sieber2023} to implement them as linear constraints.
\begin{lem}[Lemma~1 in~\cite{Sieber2023}]\label{lem:terminal-inclusion}
Consider scalars $\alpha \geq 0$, $\beta \geq 0$, matrices $A \in \mathbb{R}^{n \times n}$, $\Gamma \in \mathbb{R}^{n \times n}$, polytopic sets $\mathcal{X} = \{x \mid H_x x \leq h_x \}$ with $H_x \in \mathbb{R}^{n_x \times n}$ and $\mathcal{Y} = \{y \mid H_y y \leq h_y\}$ with $H_u \in \mathbb{R}^{n_y \times n}$, and a compact convex set $\mathcal{Z} \subset \mathbb{R}^n$, then the following conditions are sufficient and necessary for
\begin{equation}\label{eq:terminal-inclusion}
\alpha A \mathcal{X} \subseteq \beta \mathcal{Y} \ominus \Gamma \mathcal{Z}
\end{equation}
to hold:\vspace{-0.1cm}
\begin{subequations}\label{eq:terminal-lin-embed}
\begin{align}
&\exists \Lambda \in \mathbb{R}^{n_y \times n_x}, \, \Lambda \geq 0, \\
&\Lambda H_x = \alpha H_y A, \\
&\Lambda h_x \leq \beta h_y - h_{\mathcal{Z}}(\Gamma^\top H_y^\top),
\vspace{-0.1cm}
\end{align}
\end{subequations}
where $h_{\mathcal{Z}}(\Gamma^\top\! H_y^\top \!) \!=\! \begin{bmatrix} h_\mathcal{Z}(\Gamma^\top\! H_y^{{1,:}^\top}) & \mydots & h_\mathcal{Z}(\Gamma^\top\! H_y^{{n_y,:}^\top})\end{bmatrix}^\top$ denotes the stacked support functions individually evaluated at the rows of~$H_y$.
\end{lem}
\vspace{-0.5cm}
\begin{pf}
The proof can be found in~\cite[Lemma~1]{Sieber2023}.\qed
\end{pf}
\vspace{-0.4cm}
The disturbance inclusion constraints~\eqref{SLTMPC-rec-feas:dist-incl},~\eqref{SLTMPC-rec-feas:terminal-dist-incl} can both be rewritten in the form
\begin{equation*}
        \{a\} \oplus \bigoplus_{i=0}^{N-1} A_i \mathcal{X}_i \subseteq \beta \mathcal{Y} \ominus \Gamma \mathcal{Z},
\end{equation*}
by applying~\cite[Theorem~2.1]{Kolmanovsky1998} and thus can be implemented as linear constraints using the following lemma. 
\begin{lem}\label{lem:minkowski-inclusion}
Consider a scalar $\beta \geq 0$, a vector $a \in \mathbb{R}^n$, matrices $A_i \!\in\! \mathbb{R}^{n \times n}$, $\Gamma \in \mathbb{R}^{n \times n}$, polytopic sets $\mathcal{X}_i \!=\! \{x \mid H_{x,i} x \leq h_{x,i} \}$ with $H_{x,i} \in \mathbb{R}^{n_x \times n}$ and $\mathcal{Y} = \{y \mid H_y y \leq h_y\}$ with $H_u \in \mathbb{R}^{n_y \times n}$, and a compact convex set $\mathcal{Z} \subset \mathbb{R}^n$, then the following conditions are sufficient and necessary for
\begin{equation}\label{eq:minkowski-inclusion}
\{a\} \oplus \bigoplus_{i=0}^{N-1} A_i \mathcal{X}_i \subseteq \beta \mathcal{Y} \ominus \Gamma \mathcal{Z}
\end{equation}
to hold:
\begin{subequations}\label{eq:minkowski-lin-embed}
\begin{alignat}{2}
&\exists \Lambda_i \in \mathbb{R}^{n_y \times n_x}, \, \Lambda_i \geq 0, \quad &&\hspace*{-2cm}i=0, \mydots, N\!-\!1, \\
&\Lambda_i H_{x,i} = H_y A_i, \quad &&\hspace*{-2cm}i=0, \mydots, N\!-\!1,  \\
&\sum_{i=0}^{N-1} \Lambda_i h_{x,i} \leq \beta h_y - h_{\mathcal{Z}}(\Gamma^\top H_y^\top) - H_y a,
\end{alignat}
\end{subequations}
where $h_{\mathcal{Z}}(\Gamma^\top\! H_y^\top \!) \!=\! \begin{bmatrix} h_\mathcal{Z}(\Gamma^\top\! H_y^{{1,:}^\top}) & \mydots & h_\mathcal{Z}(\Gamma^\top\! H_y^{{n_y,:}^\top})\end{bmatrix}^\top$ denotes the stacked support functions individually evaluated at the rows of~$H_y$.
\end{lem}
\vspace{-0.3cm}
\begin{pf}
Using standard properties of polytopic sets, we define the scaled polytope $\beta \mathcal{Y} = \{y \mid H_y y \leq \beta h_y \}$. Then, using Theorem~2.3 in~\cite{Kolmanovsky1998}, we get
\begin{equation*}
\beta\mathcal{Y} \ominus \Gamma \mathcal{Z} = \{ y \mid H_y y \leq \beta h_y - h_{\mathcal{Z}}(\Gamma^\top \! H_y^\top) \}.
\end{equation*}
Therefore,~\eqref{eq:minkowski-inclusion} is a polytope containment problem, which can be restated as~\eqref{eq:minkowski-lin-embed} using~\cite[Prop.~1]{Sadraddini2019}.\qed
\end{pf}
\vspace{-0.3cm}
If the set $\mathcal{Y}$ in Lemma~\ref{lem:minkowski-inclusion} is restricted to a hyperrectangle, then it can be specialized to $\beta$ being a diagonal matrix, i.e., $\beta = \diag(\beta_1, \mydots, \beta_{n_y})$. This is due to the fact that for hyperrectangle $\mathcal{Y}$ and diagonal matrix $\beta$ the representation $\beta \mathcal{Y} = \{ y \mid H_y y \leq \hat{\beta} h_y \}$ with $\hat{\beta} = \diag(\beta_1 \mathbb{I}_{n}, \mydots, \beta_{n_y} \mathbb{I}_{n})$ avoids inverting $\beta$ and thus still allows optimization over $\beta$. We can use this to relax the structural constraint~$\Sigma_{i+1,i} = \sigma \cdot \mathbb{I}_n$ to $\Sigma_{i+1,i} = \textrm{diag}(\sigma_{i,1}, \mydots, \sigma_{i,n})$ as discussed in Remarks~\ref{remark:hyperrectangle} \&~\ref{remark:SLTMPC-hyperrectangle}. Additionally, if $\bar{\mathcal{W}}$ is a hyperrectangle, we recover the disturbance tube parameterization in~\cite{Chen2023} as a special case.

\subsection{Implementation of~\eqref{eq:process1}}\label{app:implementation-process1}
Since the sets $\mathcal{Z}_i$, $\mathcal{V}_i$, and $\mathcal{Q}_i^d$ in~\eqref{eq:process1} are all computed via the Pontryagin difference of a constant polytopic and an optimized polytopic set (see~\eqref{process2-output:state-input-constr},~\eqref{process2-output:dist-tube}), these sets have the same shape as the constant polytopic sets~\cite[Theorem~2.3]{Kolmanovsky1998}. For example, the sets $\mathcal{Z}^m_i = \{ z \mid H_x z \leq h_z^m \}$ all have the shape of state constraint~$\mathcal{X}$, but differ in their sizes $h_z^m$ for different memory entries. Using this observation, constraints~\eqref{process1:state_constraints} -~\eqref{process1:dist-constraint} can be implemented as linear constraints using the following lemma.
\begin{lem}\label{lem:fused-polytope}
Consider scalars $a_m \geq 0$ and polytopic sets $\mathcal{X}_m \!=\! \{x \mid H_{x} x \leq h_{x}^{m} \}$ with $H_{x} \in \mathbb{R}^{n_x \times n}$ for $m = 0, \mydots, M\!-\!1$, then the following set description holds
\begin{equation}\label{eq:fused-polytope}
\bigoplus_{m=0}^{M-1} a_m \mathcal{X}_m = \{ x \mid H_x x \leq \sum_{m=0}^{M-1} a_m h_{x}^{m} \}.
\end{equation}
\end{lem}
\begin{pf}
Using standard properties of polytopic sets, we get the scaled polytope $a_m \mathcal{X}_m = \{x \mid H_x x \leq a_m h_x^m \}$. Then, using the definition of the Minkowski sum, we get
\begin{equation*}
a_1\mathcal{X}_1 \oplus a_2\mathcal{X}_2 = \{ x_1 + x_2 \mid H_x x_1 \leq a_1 h_x^1,\, H_x x_2 \leq a_2 h_x^2 \}.
\end{equation*}
Substituting $x = x_1 + x_2$, we get the halfspace representation of $\mathcal{X} = a_1\mathcal{X}_1 \oplus a_2\mathcal{X}_2$ as
\begin{equation*}
        \mathcal{X} = \{ x \mid H_x x = H_x(x_1 + x_2) \leq a_1 h_x^1 + a_2 h_x^2 \}.
\end{equation*}
Applying this recursively for all $m = 0, \mydots, M\!-\!1$ then yields~\eqref{eq:fused-polytope}.\qed
\end{pf}
Using Lemma~\ref{lem:fused-polytope}, we can e.g. implement the inclusion constraint~\eqref{process1:state_constraints} as\vspace{-0.1cm}
\begin{equation*}
        H_x z_i \leq \sum_{m=0}^{M-1} \lambda_m h_{z,i}^{m},
        \vspace{-0.1cm}
\end{equation*}
where $h_{z,i}^m$ is the tightened state constraint stored in memory entry $m$ and computed by~\eqref{process2-output:state-input-constr} in the secondary process. In this form, the convex combination parameters $\lambda_m$ only appear in the right-hand side of the linear constraint, which allows us to optimize over them in a linear fashion.
\vspace{-0.1cm}
\section{VTOL Experiment -- Details}\label{app:rocket-details}\vspace{-0.1cm}
The dynamics of the VTOL vehicle~\eqref{eq:VTOL-dynamics} are written in matrix form for state $x = [p^x, v^x, p^z, v^z, \theta, \omega]^\top$ and input $u = [u^z, u^\theta]^\top$ as\vspace{-0.1cm}
\begin{equation*}
A = \begin{bmatrix}
1 & \Delta_t & 0 & 0 & 0 & 0 \\
0 & 1 & 0 & 0 & \Delta_t k_1 & 0 \\
0 & 0 & 1 & \Delta_t & 0 & 0 \\
0 & 0 & 0 & 1 & 0 & 0 \\
0 & 0 & 0 & 0 & 1 & \Delta_t \\
0 & 0 & 0 & 0 & \Delta_t k_2 & 1
\end{bmatrix}, \quad B = \begin{bmatrix}
0 & 0 \\
0 & 0 \\
0 & 0 \\
\Delta_t & 0 \\
0 & 0 \\
0 & \frac{\Delta_t}{I}
\end{bmatrix},
\vspace{-0.1cm}
\end{equation*}
where we use $\Delta_t = 0.075$ and $I = 0.144$ for the discretization time and inertia, respectively. The interaction parameters $k_1$, $k_2$ are assumed to lie in $[k_1^\textrm{min}, k_1^\textrm{max}] = [3.33, 4.67]$ and $[k_2^\textrm{min}, k_2^\textrm{max}] = [4.33, 5.67]$, respectively, yielding the polytopic model uncertainty description\vspace{-0.1cm}
\begin{equation*}
\Delta_A \!\in\! \textrm{co}\left\lbrace \!\begin{bmatrix} 0 & 0 & 0 & 0 & 0 & 0 \\ 0 & 0 & 0 & 0 & \Delta_t k_1^\textrm{min} & 0 \\ 0 & 0 & 0 & 0 & 0 & 0 \\ 0 & 0 & 0 & 0 & 0 & 0 \\ 0 & 0 & 0 & 0 & 0 & 0 \\ 0 & 0 & 0 & 0 & \Delta_t k_2^\textrm{min} & 0 \\ \end{bmatrix}\!, \begin{bmatrix} 0 & 0 & 0 & 0 & 0 & 0 \\ 0 & 0 & 0 & 0 & \Delta_t k_1^\textrm{max} & 0 \\ 0 & 0 & 0 & 0 & 0 & 0 \\ 0 & 0 & 0 & 0 & 0 & 0 \\ 0 & 0 & 0 & 0 & 0 & 0 \\ 0 & 0 & 0 & 0 & \Delta_t k_2^\textrm{max} & 0 \\ \end{bmatrix}\!\right\rbrace\!.
\vspace{-0.1cm}
\end{equation*}
The additive disturbance description is given by
\begin{equation*}
\mathcal{W} = \textrm{co}\left\lbrace \begin{bmatrix} 0 \\ -\sigma_w \\ 0 \\ 0 \\ 0 \\ -\sigma_w \\ \end{bmatrix}\!, \begin{bmatrix} 0 \\ \sigma_w \\ 0 \\ 0 \\ 0 \\ \sigma_w \\ \end{bmatrix}  \right\rbrace\!,
\end{equation*}
where $\sigma_w = 0.05$. The state and input constraints are then given by
\begin{align*}
\mathcal{X} \!=\! \left\lbrace\! \begin{bmatrix} -15 \\ -6 \\ 0 \\ -6 \\ -20 \\ -10 \\ \end{bmatrix}\! \leq x \leq \!\begin{bmatrix} 15 \\ 6 \\ 15 \\ 6 \\ 20 \\ 10 \\ \end{bmatrix}\! \right\rbrace \!, \ 
\mathcal{U} \!=\! \left\lbrace \!\begin{bmatrix} -5 \\ -25 \\ \end{bmatrix}\! \leq u \leq \!\begin{bmatrix} 5 \\ 25 \\ \end{bmatrix}\! \right\rbrace\!.
\end{align*}

\end{document}

%% file: tikz/asynchronous-updates.tex

\begin{tikzpicture}[>=latex,scale = 0.95]

\draw[-latex,thick] (0,-0.6) -- (5,-0.6);
\node at (0.1,-0.4) {$x$};
\draw[-latex,thick] (0.25,-0.6) -- (0.25,-1.55) -- (0.6,-1.55);

\draw[thick]  (7.4,-2.2) rectangle (5,0);
\node[align = center] at (6.25,-0.6) {Primary\\Process~\eqref{eq:process1}};

\draw[dashed]  (7.2,-2.) rectangle (5.2,-1.1);
\node[align = center] at (6.2,-1.55) {{\small Memory $\mathtt{M}$}\\{\small (\textit{Size:} $M$)}};

\draw[-latex, thick] (3.1,-1.55) -- (5,-1.55);
\node at (4.05,-1.3) {{\small $\bm{\Phi}^\mathbf{e}, \bm{\Phi}^{\bm{\nu}}$}};
\node at (4.0,-1.85) {{\small$\bm{\Sigma}, \bm{\Xi}, \alpha$}};

\draw[thick]  (3.1,-2.1) rectangle (0.6,-1);
\node[align = center] at (1.85,-1.6) {Secondary\\Process~\eqref{eq:process2}};

\draw[-latex,thick] (7.4,-0.6) -- (8.2,-0.6);
\node at (7.8,-0.4) {$u$};

\end{tikzpicture}

%% file: tikz/rocket.tex

\begin{tikzpicture}[>=latex,scale = 0.94]

\coordinate (origin) at (0,0);
\coordinate (p1) at (4,0);
\coordinate (p2) at (-4.5,0);
\coordinate (p3) at (-4.3,0);

\coordinate (rocket) at (-2,2);

\draw (p1) --++(90:4.8);
\fill[pattern=north west lines] (p1) rectangle ++(0.2,4.8);
\draw (p3) --++(90:4.8);
\fill[pattern=north west lines] (p2) rectangle ++(0.2,4.8);
\draw (p3) --++(0:8.3);
\fill[pattern=north west lines] (p2) rectangle ++(8.7,-0.2);

\draw (rocket) rectangle ++(0.4,1.75);
\node[above right, rotate=90] at (-1.55,2) {\small RockETH};
\draw (-1.8,2) -- ++(-50:0.4cm);
\draw (-1.8,2) -- ++(-130:0.4cm);

\draw[latex-latex] (-4.3,4.3) --(4,4.3);
\node[above] at (-0.15,4.3) {\small $\unit[30]{m}$};
\draw[latex-latex] (3.7,0) --(3.7,4.8);
\node[left] at (3.7,2.4) {\small $\unit[15]{m}$};

\draw[-latex] (-3.8,2.7) --(-3,2.7);
\draw[-latex] (-3.8,2.9) --(-3,2.9);
\draw[-latex] (-3.8,3.1) --(-3,3.1);
\node[above] at (-3.4,3.1) {\small wind $\unit[2]{m/s}$};

\draw[-latex] (-4.1,0.2) --(-3.5,0.2);
\draw[-latex] (-4.1,0.2) --(-4.1,0.8);
\node[above] at (-3.5,0.2) {\small $x$};
\node[right] at (-4.1,0.8) {\small $z$};

\end{tikzpicture}